\documentclass[fleqn,usenatbib]{mnras}

\usepackage[T1]{fontenc}

\DeclareRobustCommand{\VAN}[3]{#2}
\let\VANthebibliography\thebibliography
\def\thebibliography{\DeclareRobustCommand{\VAN}[3]{##3}\VANthebibliography}

\usepackage{graphicx}	
\usepackage{amsmath}	
\usepackage{amssymb}	

\title[Extragalactic HI absorption lines with FAST]{Extragalactic HI 21-cm absorption line observations with the Five-hundred-meter Aperture Spherical radio Telescope}
\author[B. Zhang et al.]{
Bo Zhang,$^{1,2}$\thanks{E-mail: zhangbo@nao.cas.cn (BZ)}
Ming Zhu,$^{1,2}$\thanks{E-mail: mz@nao.cas.cn (MZ)}
Zhong-Zu Wu,$^{3}$
Qing-Zheng Yu,$^{4}$
Peng Jiang,$^{1,2}$
You-Ling Yue,$^{1}$
\newauthor Meng-Lin Huang$^{1,2}$
and Qiao-Li Hao$^{1,2}$
\\
$^{1}$National Astronomical Observatories, Chinese Academy of Sciences, No.20A Datun Road, Chaoyang District, Beijing, 100101, China\\
$^{2}$CAS Key Laboratory of FAST, National Astronomical Observatories, Chinese Academy of Sciences\\
$^{3}$College of physics, Guizhou University, Guiyang 550025, China\\
$^{4}$Department of Astronomy, Xiamen University, Xiamen 361005, China
}

\date{Accepted XXX. Received YYY; in original form ZZZ}

\pubyear{2021}

\begin{document}
\label{firstpage}
\pagerange{\pageref{firstpage}--\pageref{lastpage}}
\maketitle

\begin{abstract}
We present a pilot study of extragalactic HI 21-cm absorption lines using the Five-hundred-meter Aperture Spherical radio Telescope (FAST). We observed 5 continuum sources with HI absorption features ﬁrstly identiﬁed in the 40\% data release of the Arecibo Legacy Fast Arecibo L-Band Feed Array (ALFA) Survey (ALFALFA), including two systems later detected by the Westerbork Synthesis Radio Telescope (WSRT). Most of our observations were carried out during the FAST commissioning phase, and we have tested different observing modes, as well as data reduction methods, to produce the best spectra. Our observations successfully conﬁrmed the existence of HI absorption lines in all these systems, including two sources that were marginally detected by ALFALFA. We ﬁtted the HI proﬁles with single or double of Gaussian functions, and calculated the HI column densities of each source. The HI absorption proﬁles obtained by FAST show much higher spectral resolution and higher S/N ratio than the existing data in the literature, thus demonstrating the power of FAST in revealing detailed structures of HI absorption lines. Our pilot observations and tests have enabled us to develop a strategy to search for HI absorption sources using the data from the FAST extragalactic HI survey, which is one of the key projects undertaken at FAST. We expect that over 1,500 extragalactic HI absorbing systems could be detected with survey data, based on sensitivity level we achieved in pilot observations.
\end{abstract}




\section{Introduction}

As the most abundant element in the Universe, hydrogen plays a vital role in the process of stellar and galaxy evolution. With its easily-traceable 21-cm hyperfine transition line, the neutral hydrogen (HI) component in galaxies provides a fundamental tool to unveil the phase transition in the interstellar medium (ISM), which is crucial for star formation activities, galaxy kinematics, as well as the large-scale cosmic structures (e.g. see \citealt{Morganti2018} and references herein). And since the strength of HI absorbing feature against radio continuum background only depends on the column density of the absorber, as well as the intrinsic properties of the background source, and is largely distance-independent (e.g., see \citealt{Carilli1998}, \citealt{Chengalur2000}, \citealt{Kanekar2003}, \citealt{Curran2006}, \citealt{Srianand2008}, \citealt{Darling2011}, \citealt{Allison2012},  \citealt{Morganti2015}, \citealt{Wu2015}, \citealt{Allison2016},  \citealt{Maccagni2017} and Song et al. 2021, in preparation), compared with flux-limited and telescope sensitivity-restricted emission line observations, HI absorbers provide a chance to uncover the HI content in the high redshift universe (e.g., see \citealt{Kanekar2004}, \citealt{Morganti2015}, \citealt{Allison2016}, \citealt{Curran2019}, and references herein), with samples of $z>3$ absorbers detected (\citealt{Uson1991}, \citealt{Kanekar2007}). Also, such HI absorption lines can bring insights into the continuum sources, usually active galaxy nuclei (e.g. see \citealt{Maccagni2017}, \citealt{Morganti2018} and references herein), thus revealing possible interactions between star formation and central supermassive black hole activities.

However, only chance alignments between radio-loud active galactic nuclei (AGNs) and foreground or associated neutral hydrogen gas can give rise to HI absorption lines \citep{Morganti2018}, and sufficiently high column density or large mass of absorbing HI is required for the production of prominent absorption lines \citep{Darling2011}, because of the low Einstein rate of HI emission. Thus, theoretically speaking, due to the occasional nature of HI absorption lines, extensive observations along numerous line of sights are required in order to accumulate a larger sample of HI absorbing systems \citep{Darling2011}. A more practical way to search for HI absorption features is to perform radio observations on damped Lyman-$\alpha$ (DLA) systems, which features high HI column density. However, since the comoving density of low-redshift ($z \leqslant$ 0.1) DLA systems is relatively low \citep{Zwaan2005}, such efforts bear limited results (e.g., see \citealt{Kanekar2009}). 

On the other hand, the absorption line search from AT20G compact radio galaxies using the Australia Telescope Compact Array (ATCA, \citealt{Allison2012}), the 21 cm Spectral Line Observations of Neutral Gas with the EVLA (21-SPONGE) survey \citep{Murray2015}, the Westerbork Synthesis Radio Telescope (WSRT) radio galaxy surveys (\citealt{Gereb2014}, \citealt{Gereb2015}, \citealt{Maccagni2017}), as well as searches by the Australian SKA Pathfinder (ASKAP, \citealt{Glowacki2019}, \citealt{Sadler2020}), all targeting known quasars with radio interferometers, have identified dozens of extragalactic HI absorption lines. And similar attempts has been made with the Millennium Arecibo 21 cm Absorption-Line Survey (\citealt{Heiles2003a}, \citealt{Heiles2003b}), although using a single-dish telescope. Such targeted observations are mainly focused on extragalactic radio sources with sufficiently high fluxes, thus can only lead to continuum flux-biased HI absorber samples, with galaxies exhibiting weak radio emissions while high HI column densities, which can also give rise to noticeable absorption features, largely missing (e.g., see \citealt{Sadler2007}).

``Blind'' searches through HI surveys performed by large aperture single-dish telescopes provide chances to uncover more HI absorption systems. Although such surveys usually utilise the non-tracking, drift scan observing strategy, with integration time for each individual source is limited, the collecting areas of the participating telescopes can still achieve considerable sensitivities. \cite{Allison2014} identified 4 HI absorbers, including one previously unknown source, with the archival data of the HI Parkes All-Sky Survey (HIPASS, see \citealt{Barnes2001}), while \cite{Darling2011}, \cite{Wu2015}, as well as Song et al. (2021, in preparation) has made attempts to perform ``blind'' searches for absorption features using part of Arecibo Legacy Fast Arecibo L-Band Feed Array (ALFA) Survey (ALFALFA, see \citealt{Giovanelli2005}, \citealt{Haynes2018}) data, respectively, with ten sources identified in total, including 3 samples remained undetected by other instruments so far, that is, UGC 00613, CGCG 049-033 and PGC 070403, thus proving the feasibility of searching for new HI absorbers with massive blind sky surveys. Compared with radio interferometers, large single-dish telescopes such as Arecibo can provide better sensitivities, and usually higher spectral resolution, which are all crucial to reveal the characteristics of extragalactic HI lines. And although the spatial resolution for such instruments are quite limited compared with interferometers, the chance of having two or more HI absorbing systems lying within the beam width with similar redshift is quite low, thus making confusions in source identification unlikely to happen. However, it should be noted that due to temporal variations in spectral baseline commonly seen during drift scans, follow-up observations are  needed for reliable characterisation of the newly identified absorbers, especially for the weak ones. Also, interferometric mappings are usually required to discern possible fine structures within each absorbing system. 

The newly commissioned Five-hundred-meter Aperture Spherical radio Telescope (FAST) \citep{Nan2006, Nan2011, Jiang2019} is the largest filled-aperture single-dish radio antenna in the world, with its sensitivity and observable sky coverage both surpassing those of the Arecibo dish. Thus, it is nature to expect better HI absorption observations to be obtained with FAST. Compared with $\sim$several HI absorber discoveries made with existing blind surveys by \cite{Wu2015}, Song et al. (2021, in preparation), and \cite{Allison2014}, \cite{Wu2015} predicted that the number of absorbing systems detected by FAST should achieve an order of magnitude of at least $10^2$ with the upcoming Commensal Radio Astronomy FasT Survey (CRAFTS, see \citealt{Li2018}), and \cite{Yu2017} gave an expectation of over 1,000 extragalactic HI absorption line detections through blind searches in FAST drift scan data. This means that the number of known HI absorbers could be increased by 10 times with FAST, considering currently only $\geqslant 100$ such systems with redshift $z<1$ exist \citep{Chowdhury2020}. And the FAST observations will also complement the ongoing HI absorption surveys performed by the next generation arrays, including the MeerKAT Absorption Line Survey (MALS, see \citealt{Gupta2016}), the Widefield ASKAP L-band Legacy All-sky Blind surveY (WALLABY, see \citealt{Koribalski2020}), the First Large Absorption Survey in HI (FLASH) by ASKAP (\citealt{Allison2020}), the Search for HI Absorption with AperTIF (SHARP) survey (\citealt{Adams2018}), as well as the uGMRT Absorption Line Survey (\citealt{Gupta2021}), with a more complete Northern Sky coverage, better spectral resolution compared with all these surveys, and a similar noise level obtained during much shorter integration times provided by FAST.

In this paper, we report extragalactic HI absorption line observations conducted with FAST on five galaxies identified by \cite{Wu2015} and Song et al. (2021, in preparation), as a pilot study of massive HI absorption line observations by FAST in the near future. The content of this paper is organised as follows. In Section \ref{sec:2}, we introduce the galaxy samples, FAST observation settings, as well as the data reduction process in details. The related results are presented and summarised in Section \ref{sec:3}. In Section \ref{sec:4} we give discussions, and conclusions are drawn in Section \ref{sec:5}. Here, we adopt the $\Lambda$CDM cosmology, assuming the cosmological constants as $H_0 = 70$ km s$^{-1}$ Mpc$^{-1}$, $\Omega_M = 0.3$, and $\Omega_{\Lambda} = 0.7$. And in accordance with the ALFALFA survey \citep{Haynes2018}, the optical definition of line speed ($\delta \lambda/\lambda$) rather than the radio one ($\delta \nu/\nu$) is applied throughout our data reduction process.

\section{Sample selection and FAST observations}\label{sec:2}
\subsection{Sample selection}

As a pilot study, we tested various observing modes and adopted different backend set-ups to observe the HI spectra during the FAST commissioning phase. The HI absorption systems are particularly suitable for such a purpose, as they contain both continuum sources and spectra lines with know fluxes, thus providing good opportunities for checking telescope pointing accuracy and testing our calibration methods. We selecte 5 absorbing systems firstly identified from 40\% of ALFALFA data by \cite{Wu2015}, including UGC 00613, ASK 378291.0, CGCG 049-033, J1534+2513, and PGC 070403. Among them, ASK 378291.0 and J1534+2513 were also reported by the WSRT survey for radio galaxies described by \cite{Maccagni2017}, with J1534+2513 and PGC 070403 already marked as possible absorptions by \citep{Haynes2011}. For each system in our samples, the continuum background is provided by a previously-detected radio source, while the absorbing HI gas lies within the source itself. The basic information of these sources are listed in Table \ref{tab1}.

\begin{table*}
	\centering
	\renewcommand\arraystretch{1.5}
	\caption{Basic characteristics of the 5 targets observed by FAST.}
	\label{tab1}
		\begin{tabular}{llccccccc} 
		\hline
		Source name & AGC no.$^{\left(a\right)}$ & RA (J2000)$^{\left(b\right)}$ & Dec (J2000)$^{\left(b\right)}$ & $cz$  (km s$^{-1}$)$^{\left(c\right)}$ &  $S_{ 1.4{\rm GHz, NVSS} }$  (mJy)$^{\left(d\right)}$ &  $S_{1.4{\rm GHz, FIRST}}$ (mJy/beam)$^{\left(e\right)}$\\
		\hline
		UGC 00613 & 613 & 00h 59m 24.42s & +27d 03m 32.6s & $13770.07 \pm 23.08$ & {$112.6 \pm 4.0$} & - &  \\
        ASK 378291.0 & 712921 & 10h 25m 44.23s & +10d 22m 30.5s & $13732.00 \pm 26.98$ & $76.6 \pm 2.3$ & $73.85 \pm 0.145$ \\
        CGCG 049-033 & 250239 & 15h 11m 31.38s & +07d 15m 07.1s & $13383.94 \pm 35.98$ & $108.8 \pm 3.8$ & $79.50 \pm 0.151$ \\
        J1534+2513 & 727172 & 15h 34m 37.62s & +25d 13m 11.4s & $10180.65 \pm 2.698$ & $50.1\pm 1.6$ &$44.58 \pm 0.146$ \\
        PGC 070403 & 331756 & 23h 04m 28.24s & +27d 21m 26.5s & $7520.894 \pm 118.4$ & $116.6\pm 3.5$ & - \\
		\hline
	\end{tabular}
	
$^{\left(a\right)}$ Serial number in the Arecibo General Catalogue (AGC), see \cite{Haynes2018}.

$^{\left(b\right)}$ Equatorial coordinates for corresponding optical counterparts retrieved from NASA/IPAC Extragalactic Database (NED). 

$^{\left(c\right)}$ Heliocentric redshift data retrieved from NED. 

$^{\left(d\right)}$ 1.4 GHz integrated flux data retrieved from the NRAO Very Large Array (VLA) Sky Survey (NVSS) catalogue \citep{Condon1998}.

$^{\left(e\right)}$ 1.4 GHz peak flux data retrieved from the VLA Faint Images of the Sky at Twenty-Centimeters (FIRST) survey \citep{Becker1995}. (For comparison only in order to keep consistency with \citealt{Wu2015} and Song et al. 2021, in preparation; not used in data reduction.)
\end{table*}

\subsection{Observation setting-ups}

Four of the five sources have been observed with the tracking mode during the commissioning phase of the FAST telescope. And two of these, ASK 378291.0 and UGC 00613, have been observed on October 29 and November 5 of 2017, respectively, with the wide band receiver covering the $270 - 1620$ MHz frequency range. Since the spectral backend for FAST was still in development at that time, an off-the-shelf N9020A MXA spectrum analyser produced by Keysight Technologies. Inc was adopted as the primary spectrometer, with an integration time of $\sim 6$ s for each spectrum, and 1001 frequency channels over a 6 MHz band, thus achieving a spectral resolution of $\sim 1.2$ km s$^{-1}$, which brought significant improvements towards ALFALFA's 5.3 km s$^{-1}$ \citep{Giovanelli2005}, or WSRT survey's 16 km s$^{-1}$ \citep{Maccagni2017}. Data from each source was recorded for 1200 s, with signals from one polarisation only being recorded. According to the optical redshifts of the sources, the frequency coverage of the spectrum analyser was chosen as $1354.0 - 1360.0$ MHz for ASK 378291.0, and $1354.3- 1360.3 $ MHz for UGC 00613. 

\begin{figure}
\includegraphics[width=\columnwidth]{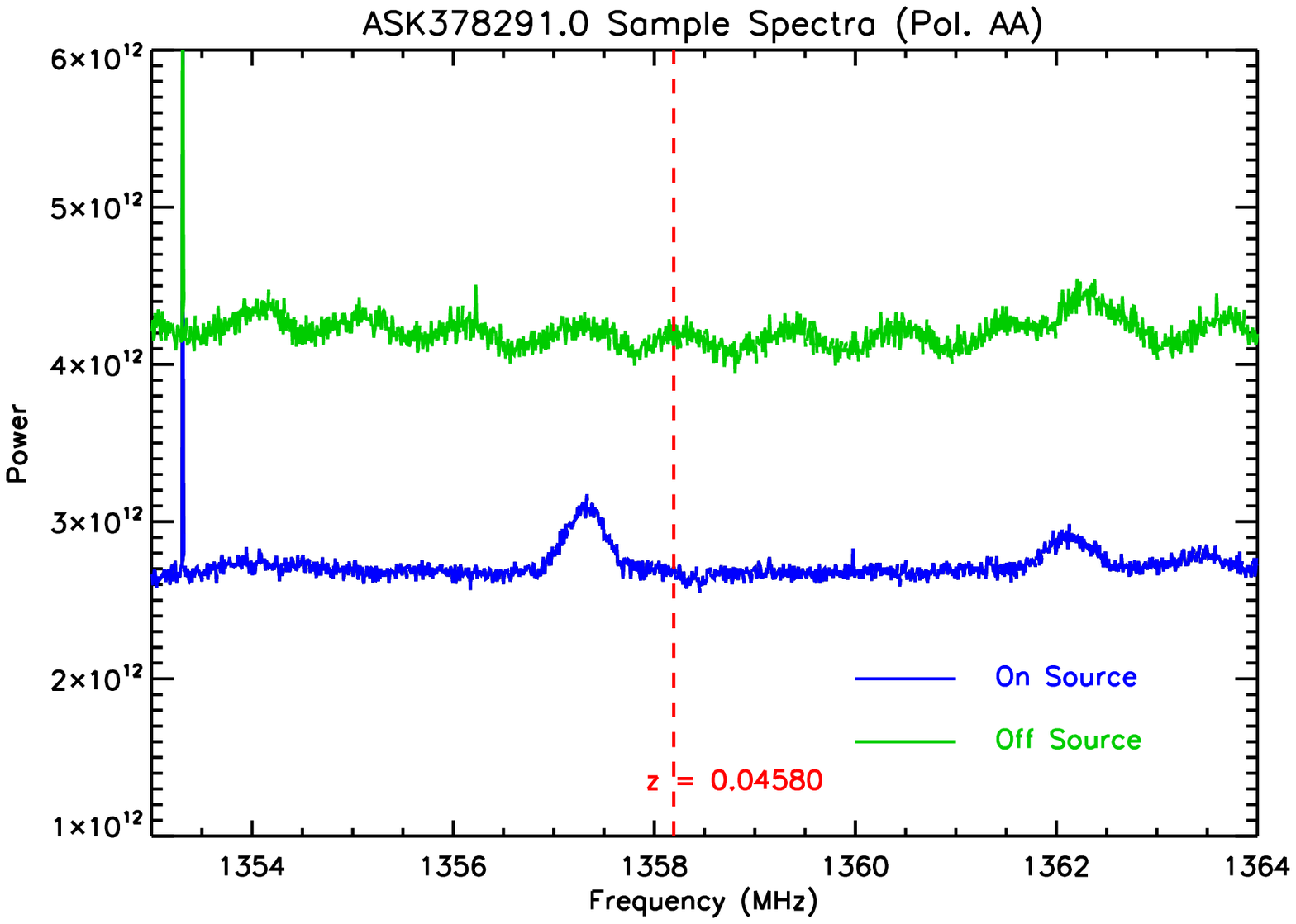}\\
\includegraphics[width=\columnwidth]{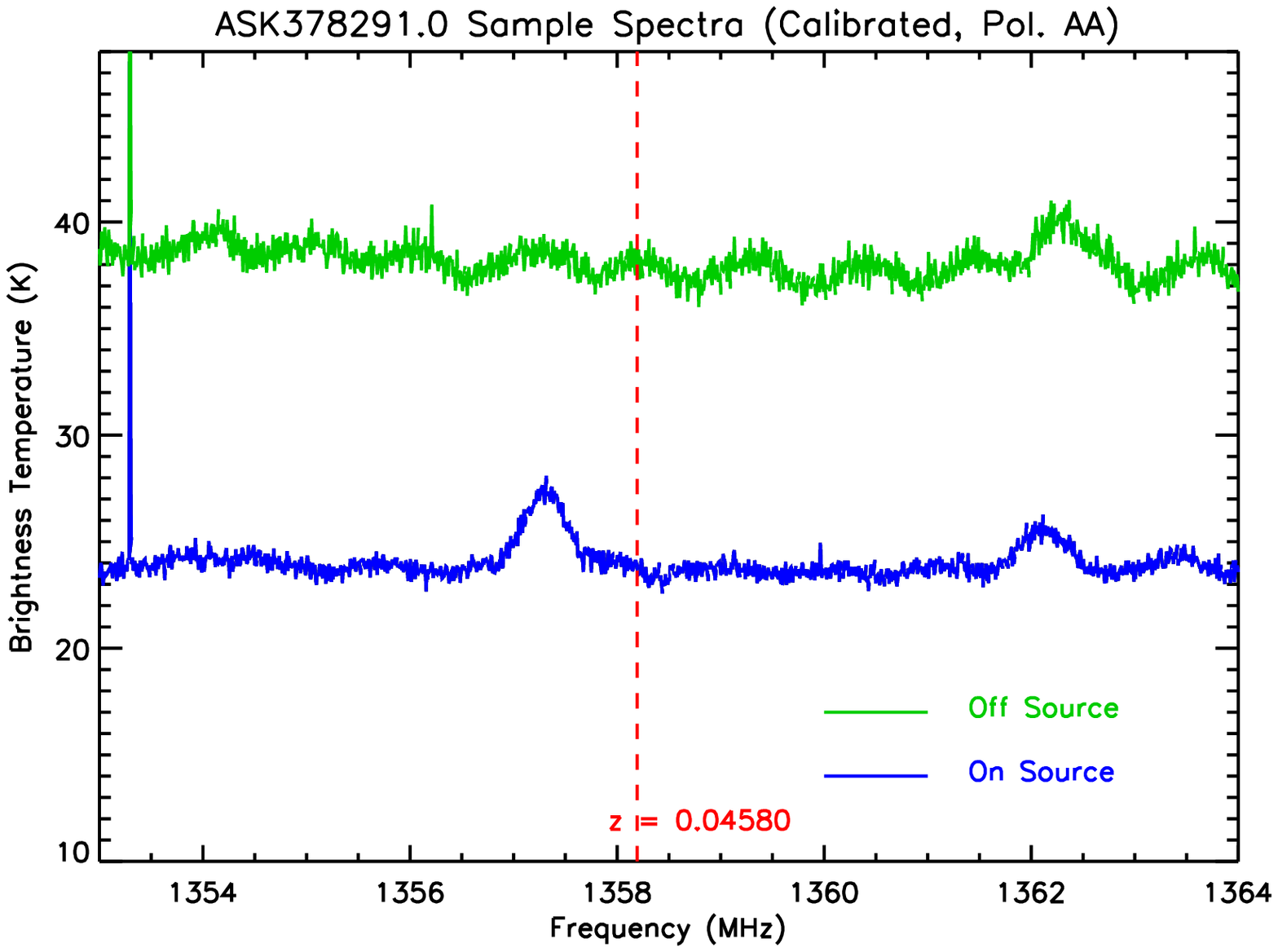}
\caption{Top: The original, unprocessed on-source (blue) and off-source (green) data sample of ASK 378291.0. Bottom: Noise diode-calibrated on-source (blue) and off-source (green) data of ASK 378291.0. It can be seen that the flux levels and standing wave behaviours of the two spectra differ from each other significantly, even for the calibrated data. Besides, the off-source data exhibits a higher background level, which is possibly due to temporal instrumental baseline fluctuations, as well as broad-band RFI contamination. The red dotted line mark the frequency of the desired absorption feature. Observed frequencies shown in this figure are not heliocentric-corrected.} 
\label{fig1}
\end{figure}

With the 19-beam L-band receiver covering the band of $1050 - 1450$ MHz \citep{Smith2017} and its spectral backend inaugurated in the summer of 2018, we were able to observe sources ASK 378291.0, CGCG 049-033, J1534+2513, and PGC 070403 with this new set of instruments. The first three of these 4 sources were observed using the tracking mode, with data from the central beam (Beam 01) only being recorded. The 19-beam spectral backend can take data from both polarisations with 65,536 spectral channels covering the $1000-1500$ MHz band (equivalent to $\sim 1.5$ km s$^{-1}$ spectral resolution), and the typical integration time for each spectrum is $\sim 1$ s \citep{Jiang2020}. ASK 378291.0 was observed for 382.5 seconds on September 2, 2018, while data for CGCG 049-033 and J1534+2513 were recorded for 1200 s each on October 31, 2018. And an extra round of tracking observation for CGCG 049-033 was performed on December 28, 2018, also lasting 1200 s. The last source, PGC 070403 with a relatively weak absorption feature as reported by \cite{Wu2015} and Song et al. (2021, in preparation), was observed as a by-product of the drift scan conducted on September 14, 2020. One of the aims of this observation, during which the footprint of Beam 01 passed directly over PGC 070403 at the transit time of this target, was to verify the feasibility of finding extragalactic HI absorbing systems with the upcoming CRAFTS survey. The detailed of each observing session are listed in Table \ref{setup}.

\begin{table*}
	\centering
	\renewcommand\arraystretch{1.5}
	\caption{FAST observation setting-ups for the 5 targets.}
	\label{setup}
		\begin{tabular}{llcccclcc} 
		\hline
		  Source name & Obs. Mode & Obs. Session & Receiver$^{\left(a\right)}$ & Backend$^{\left(b\right)}$ & Band Coverage & Channel No. & Obs. Date & Duration \\
              & &  & &     & (MHz) &  & & (s) \\\hline
  UGC 00613 & Tracking & 1 & W & A & $1354.3 - 1360.3$ & 1,001 & 2017 Nov. 05 & 1200\\
  \hline
  ASK 378291.0 & Tracking & 1 & W & A & $1354 - 1360$ & 1,001 & 2017 Oct. 29 & 1200\\
               & Tracking & 2 & 19 & S & $1000-1500$ & 65,536 & 2018 Sep. 02 & 382.5\\
  \hline
  CGCG 049-033 & Tracking & 1 & 19 & S & $1000-1500$ & 65,536 & 2018 Oct. 31 & 1200\\
               & Tracking & 2 & 19 & S & $1000-1500$ & 65,536 & 2018 Dec. 28 & 1200\\
  \hline
  J1534+2513   & Tracking & 1 & 19 & S & $1000-1500$ & 65,536 & 2018 Oct. 31 & 1200\\
  \hline
  PGC 070403   & Drifting & 1 & 19 & S & $1000-1500$ & 65,536 & 2020 Sep. 14 & $\sim 12$\\
  \hline
		\hline
	\end{tabular}
	
$^{\left(a\right)}$ ``W'' for the wide band receiver; ``19'' for the 19-beam receiver.

$^{\left(b\right)}$ ``A'' for Keysight spectrum analyser; ``S'' for spectral line backend.
\end{table*}

In order to reduce the effects of baseline fluctuations, we chose an ``off-source'' position in the sky for each source observed with the tracking mode, which is located several arcminutes away from the primary target, as our reference for background. However, it should be noted that as seen in Fig. \ref{fig1}, due to phase-shifting behaviours of standing waves and instabilities of instruments commonly seen during the FAST commissioning phase, as well as time-dependent broad-band radio frequency interference (RFI), the continuum flux level obtained by comparing on- and off-source positions fluctuates significantly between different observing sessions, even for the same source. And cases exhibiting higher off-source baseline level flux than on-source spectra, as well as curved bandpass exist, which makes a direct measurement of continuum level by FAST unreliable. Thus, in the following data reduction process, we mainly extract the background ``baseline'' by directly fitting background continuum with on-source data, rather than utilising the off-source observations. And the details of our data reduction process is described in the following subsection.

\subsection{Flux calibration and baseline subtraction}

\begin{figure}
\includegraphics[width=\columnwidth]{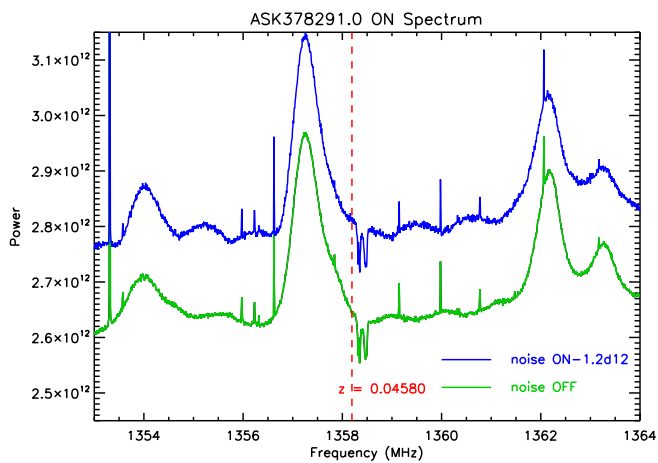}
\caption{Averaged on-source spectra of ASK 378291.0 with noise ON (blue, offset in y-axis added for convenience) and OFF (green). The absorption feature can be found near to $\sim 1359$ MHz, indicated by the red dotted line. It can be seen that the phase of standing waves imposed on the background continuum flips between the ON and OFF states. Frequencies shown in this figure have not been heliocentric corrected yet.} 
\label{fig2}
\end{figure}

The 19-beam receiver of FAST is equipped with built-in noise diode for signal calibration. As illustrated in \cite{Jiang2020}, the diode can be operated at 2 modes, with characteristic noise temperatures $T_{noise}$ as $\sim 1.2$ K and $\sim 12$ K, respectively. For ASK 378291.0, CGCG 049-033, J1534+2513, and PGC 070403 observations, we all adopted the high-temperature mode of the noise diode, and the calibration signal was injected periodically during each observing session. Let ${\rm OFF}$ be the original instrument reading without noise, ${\rm ON}$ the reading with noise injected (it should be noted here that throughout this work, we adopt capitalised ON and OFF to describe the noise diode status, while the lower case on/off denote the on/off source position).  $T_{noise}$ the pre-determined noise temperature measured with hot loads, which has been shown in \cite{Jiang2020}, the system temperature $T_{sys}$ can be computed as
\begin{equation}
T_{sys} = \frac{{\rm OFF}}{{\rm ON} - {\rm OFF}} \times T_{noise}.
\label{eq:1}
\end{equation}
For each target observed with the 19-beam receiver, we calculated averaged $T_{sys}$ with Eq. \ref{eq:1} using averages of ${\rm OFF}$ and ${\rm ON}$ during each observing session. And as seen in Fig. \ref{fig2}, since the phase of instrumental standing waves changes with each activation of the high-temperature noise, and the wavelength of the standing wave ripples induced by the FAST receiving system can be approximated as $ c/2f \sim 1.1$ MHz (where $c$ is the speed of light, and $f \sim 138$ m the focal length of FAST), corresponding to $\sim 140-150$ spectral channels, we performed a several-hundred-channel Gaussian smooth on each set of ${{\rm ON} - {\rm OFF}}$ value, to minimise the phase-shifting effects in calibrations. Then $T_{sys}$ from both polarisations were manually checked, and if they were consistent with each other (which was the case for all four samples observed by the 19-beam receiver), each pair of polarisation A \& B temperature readings were added together to get the average. Finally, the polarisation-averaged $T_{sys}$ was converted to flux densities with pre-measured antenna gain ($\sim 15.7- 16.5$ K/Jy for Beam 01, depending on frequency, see Table 5 in \citealt{Jiang2020} for details).  

However, during the ``early-science'' stage of FAST's commissioning phase, when the wide band receiver was still in use, no reliable noise diode was available for flux calibrations. What was worse, the level of instrument reading for N9020A MXA spectrum analyser is often relatively unstable, even within one observing session. Thus, it poses significant difficulties in calibrating the UGC 00613 data. Fortunately, the absorption feature of ASK 378291.0 was detected with both the wide band receiver and the 19-beam array. Thus, we adopted the measured flux density of ASK 378291.0 from the latter instrument as reference to re-scale and calibrate the wide band receiver observations. And since only one polarisation was available during this period, we just assumed that the single set of flux densities applies for both polarisations. Although such operations could bring considerable compromise to the accuracy of our final products, it is the only sensible method that could  produce reasonable results generally compatible with the ALFALFA observations, as provided by \cite{Wu2015} and Song et al. (2021, in preparation). More details of our results can be found in Section 3.1.1.

Once the flux densities were calculated via calibration, the observed frequencies were Doppler-corrected to the heliocentric frame. Since the fully automatic data reduction pipeline for FAST HI observations \citep{Zhang2019} is still in development, all baseline subtraction and RFI flagging works involved in this paper were performed manually. For every source, the overall shape of the background continuum (``baseline'') curve was determined by calculating the median value for every channel during each entire session, similar to the process utilised by HI Parkes All Sky Survey (HIPASS, see \citealt{Barnes2001}). As for the baseline reading for the frequency channels in which the desired HI absorption lines resides, a cubic spline interpolation was performed. Once the baseline was subtracted, strong RFI was marked by visual inspections. Finally, 1- or 2-component Gaussian fitting was applied to data not contaminated by RFI, depending on the line profiles, to get the spectral parameters. Here considering the baseline instabilities between on- and off-source positions as described in Subsection 2.2, continuum flux densities measured by the NVSS survey (rather than direct measurements by FAST) were adopted for the calculation of optical depth and other line characteristics of each HI absorbing feature. 

Here it should be noted that since the NVSS survey was performed more than 2 decades ago, it is quite possible that the continuum flux level for the observed sources varies during this period. In fact, as shown in Table \ref{tab1}, the continuum flux of CGCG 049-03 does show a difference as large as 27\% between NVSS and the earlier VLA Faint Images of the Sky at Twenty-Centimeters (FIRST) survey \citep{Becker1995} results, which may be due to possible source variability, or the existence of the extended jet in this source (\citealt{Bagchi2007}), combined with the varied VLA configurations adopted by different surveys. (While for ASK 378291.0 and J1534+2513, which have also been observed by both NVSS and FIRST, measurements obtained by these two surveys are generally compatible, with flux variations less than $\sim10$\%.) Also, the beam size difference between NVSS survey and FAST, along with the unfilled $u-v$ coverage of VLA may cause the NVSS flux to be somewhat underestimated for this work. Thus, as mentioned in Section 2.1, our strategy to calculate absorption parameters (including HI column density $N_{I}$ and optical depth $\tau$) based on existing survey results is only a compromise to the not-so-stable behaviours of our newly commissioned instrument, and is adopted in reference to the solution utilised by various single-dish absorption line observations, such as \cite{Darling2011}, \cite{Grasha2019}, as well as \cite{Zheng2020}.

\section{Observing results}\label{sec:3}

We successfully detected HI absorption features in all of the 5 sources firstly identified with 40\% of the ALFALFA data by \cite{Wu2015} and Song et al. (2021, in preparation). The properties of each detected source estimated using FAST are summarised in Table \ref{tab2}. The column densities of HI towards absorption line $N_{HI}$ in this table are calculated with the following equation 
\begin{equation} 
N_{HI} = 1.823 \times 10^{18} \frac{T_s}{f} \int \tau {\rm d} v \text{ cm}^{-2}. 
\label{eq:2} 
\end{equation}
Here $T_s$ denotes the HI spin temperature, $f$ the covering factor of the radio continuum source, with a typical value of unity assumed (e.g., see \citealt{Maccagni2017}). Since no source observed by this work is associated with known DLA system \citep{Wu2015}, which usually exhibits a higher $T_s$ in the order of $\sim 10^3$ K {(e.g., see \citealt{Chengalur2000}, \citealt{Darling2011})}, and \cite{Curran2019}, a typical value of $T_s\sim 100$ K is assumed. And the optical depth $\tau$ of the HI absorber can be calculated as 
\begin{equation}
\tau = -\ln\left(1+\frac{S_{HI}}{S_{1400 \text{ MHz}}} \right),
\label{eq:3}
\end{equation}
where $S_{HI}$ is the depth of the absorption line (in negative value), and $S_{1.4 \text{ GHz}}$ the 1400 MHz flux of the continuum source as provided by NVSS survey. All errors listed in Table \ref{tab2} have been evaluated with the spectral resolution and RMS level of each observing session, the calibration accuracy of the 19-beam receiving system ($\sigma_{cal} \sim 0.8-1.8$\%, as noted by \citealt{Jiang2020}), the $\sigma_{S_{1.4{\rm GHz}}}$ value of NVSS or VLA FIRST measurements, as well as the line profiles, in reference to the method adopted by \cite{Koribalski2004}, as follows

\begin{equation}
\sigma_{ cz_{peak} }  =  3\frac{\sqrt{P \Delta_v}}{{\rm S/N}},
\end{equation}
\begin{equation}
\sigma_{{\rm FWHM}}  =  2 \sigma_{cz_{peak}},
\end{equation}
\begin{equation}
\sigma_{S_{HI,peak}}  =  \sqrt{ {\rm RMS}^2 + \sigma_{cal}^2\times S_{HI,peak}^2},
\end{equation}
where $S_{HI,peak}$ the maximum line depth, $P = 0.5\times({\rm FW}20-{\rm FWHM})$ the slope of the HI line profile, ${\rm FW}20$ the line width measured at 20\% of the peak flux value, FWHM the full width at half maximum of the line, $\Delta_v$ the spectral resolution in km s$^{-1}$, ${\rm S/N} = -S_{HI,peak} / {\rm RMS}$ the signal-to-noise ratio, $cz_{peak}$ the corresponding redshift of the line peak, and RMS the root mean square of the background noise extracted from both side of the absorption line. Thus, the error of optical depth at the line peak can be evaluated with Eq. \ref{eq:2}, using the error propagation theory, as
\begin{equation}
\sigma_{\tau_{peak}}  =  \sqrt{\left(\frac{\partial \tau}{\partial S_{HI,peak}} \sigma_{S_{HI,peak}}\right)^2 + \left(\frac{\partial \tau}{\partial S_{1.4 \text{ GHz}}} \sigma_{S_{1.4 \text{ GHz}}} \right)^2 }.
\end{equation}
And considering $N_{HI} \approx 1.823 \times 10^{18} \left(T_s/100 {\rm K}\right) \times 1.064 \cdot {\rm FWHM} \cdot \tau_{peak}$ cm$^{-3}$ for Gaussian line profiles, as shown by \cite{Darling2011, Murray2015}, the error of HI column density  $\sigma_{ N_{HI} }$ can be estimated as
\begin{equation}
\sigma_{ N_{HI} }  =  \frac{ 1.940\times 10^{18} T_s}{100 {\rm K}} \sqrt{ \left(\tau_{peak} \cdot \sigma_{{\rm FWHM}} \right)^2 + \left( {\rm FWHM} \cdot \sigma_{\tau_{peak}}\right)^2} .
\end{equation}

\begin{table*}[th]
\caption{FAST observations of the five extragalactic HI absorbers firstly identified with ALFALFA data. All parameters are listed in the rest frame of the absorbers. All background RMS data are measured with the original velocity bin. Parameters including $cz_{peak}$, FWHM, $S_{HI,peak}$, $\tau_{peak}$, and $\int \tau {\rm d} v$ are calculated with the fitted line profiles.}\label{tab2}
\begin{tabular}{llccccccc}
  \hline
  Source name & RMS & Gaussian & $cz_{peak}$ & FWHM       & $S_{HI,peak}$ &$\tau_{peak}$ & $\int \tau {\rm d} v$ & $N_{HI}$\\
              &(mJy)& Component&(km s$^{-1}$)& (km s$^{-1}$)&  (mJy)    &           & (km s$^{-1}$)         & ($10^{20}$ cm$^{-2}$)\\\hline
  UGC 00613 & 6.49 & A & $13956.5\pm 0.9$ & $26.36\pm1.74$ & $-64.28\pm 6.61$ & $0.801\pm 0.145$ & $21.24\pm 4.10$ & $38.72 \pm 7.95\left( \frac{T_s}{100\text{ K}} \right)$ \\
  \hline
  ASK 378291.0 & 0.89 & A & $13672.3\pm0.2$ & $14.06\pm 0.36$ & $-42.24\pm 1.22$ & $0.607 \pm 0.051$ & $9.095 \pm 0.753$ & $16.58 \pm 1.46 \left(\frac{T_s}{100\text{ K}} \right)$\\
               &       & B & $13702.2\pm0.1$ & $14.09\pm 0.30$ & $-45.93 \pm 1.28$ & $0.683 \pm 0.061$ & $9.724 \pm 0.887$ & $17.72 \pm 1.72\left(\frac{T_s}{100\text{ K}} \right)$\\
               &       & Total &$13687.3\pm0.1$& $44.00\pm 0.30$ &     &       & $18.83\pm 1.16$ & $34.33 \pm 2.26\left( \frac{T_s}{100\text{ K}} \right)$\\ 
  \hline
  CGCG 049-033 & 0.45 & A & $13454.4\pm 1.0$ & $52.71\pm2.07$ & $-5.851 \pm 0.468$ & $0.055\pm 0.005$ & $3.069 \pm  0.285$ & $5.595 \pm 0.553\left(\frac{T_s}{100\text{ K}} \right)$\\ 
  \hline
  J1534+2513   & 0.37 & A & $10089.0 \pm 0.1$ & $5.148 \pm 0.240$ & $-14.97 \pm 0.48$ & $0.355 \pm 0.019$ & $2.291 \pm 0.130$ & $4.176 \pm 0.253 \left( \frac{T_s}{100\text{ K}} \right)$\\
               &       & B & $10107.9 \pm 0.2$& $13.76 \pm 0.33$ & $-15.32 \pm 0.48$ & $0.365 \pm 0.020$ & $4.713 \pm 0.296$ & $8.593 \pm 0.575\left(\frac{T_s}{100\text{ K}} \right)$\\
               &       &Total&$10098.5\pm 0.1$& $27.53 \pm 0.29$ &          &       & $7.011 \pm 0.324$ & $12.78 \pm 0.63\left(\frac{T_s}{100\text{ K}} \right)$\\
  \hline
  PGC 070403   & 2.53 & A & $7694.2\pm 4.2$ & $74.54\pm 8.37$ & $-9.602\pm 2.539$ & $0.086\pm 0.024$ & $6.643 \pm 2.314$ & $12.11 \pm 4.49\left(\frac{T_s}{100\text{ K}}\right)$\\
  \hline
\end{tabular} 
\end{table*}

\subsection{Note on individual sources and comparisons with Arecibo/WSRT observations}

\subsubsection{ASK 378291.0}

\begin{figure}
\includegraphics[width=\columnwidth]{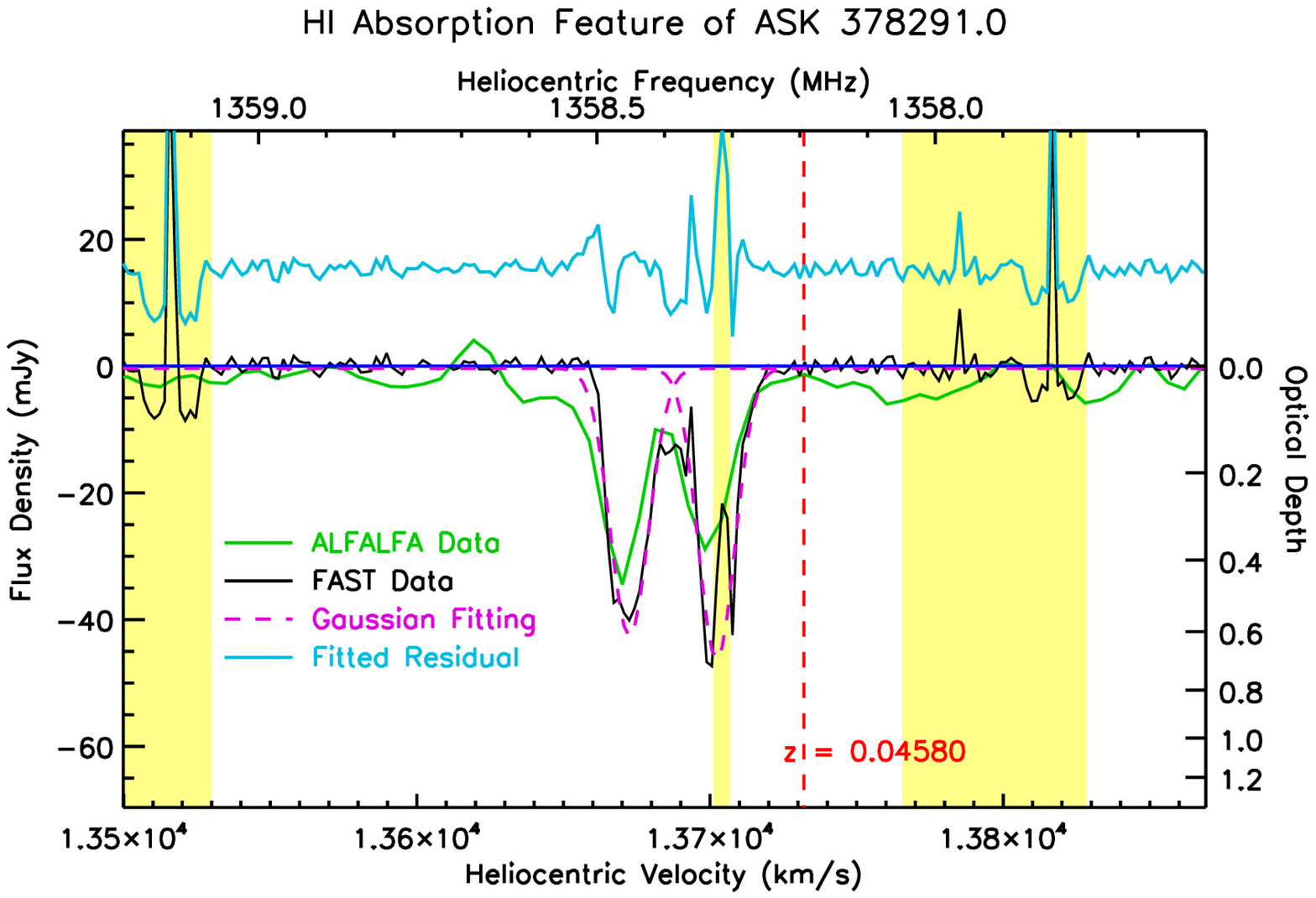}\\
\includegraphics[width=\columnwidth]{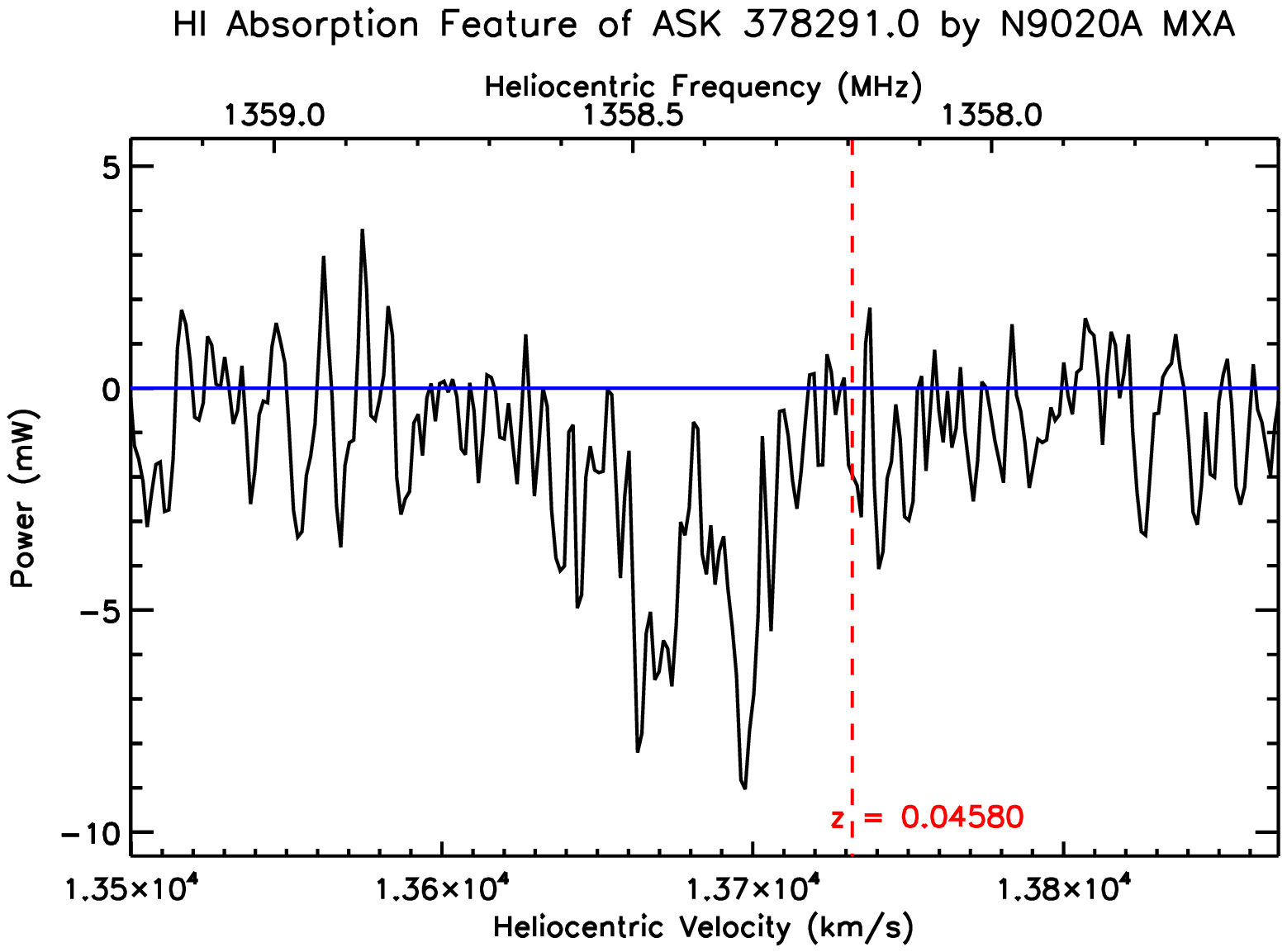}
\caption{Top: HI absorption feature ASK 378291.0. The black line shows observations taken by the FAST 19-beam receiver with its spectral line backend on Sep. 2, 2018, while the green curve denotes data taken from the ALFALFA survey. The magenta dotted line is the fitting result by dual-Gaussian functions, while the cyan curve shows the fitting residual with a 15 mJy offset for clarity. The regions shaded with light yellow are manually-flagged RFI, which have been all excluded in the fitting procedure and RMS computations. The red dotted line shows the optical redshift, while the blue line marks the zero flux level. Bottom: The same target observed by the wide band receiver along with the N9020A MXA spectrum analyser on Oct. 29, 2017, with the original, uncalibrated instrumental readings measured in mW shown along the y-axis. This set of data marks the first successful detection of extragalactic HI absorption line by the FAST telescope.} 
\label{figASK}
\end{figure}

\begin{figure}
\includegraphics[width=\columnwidth]{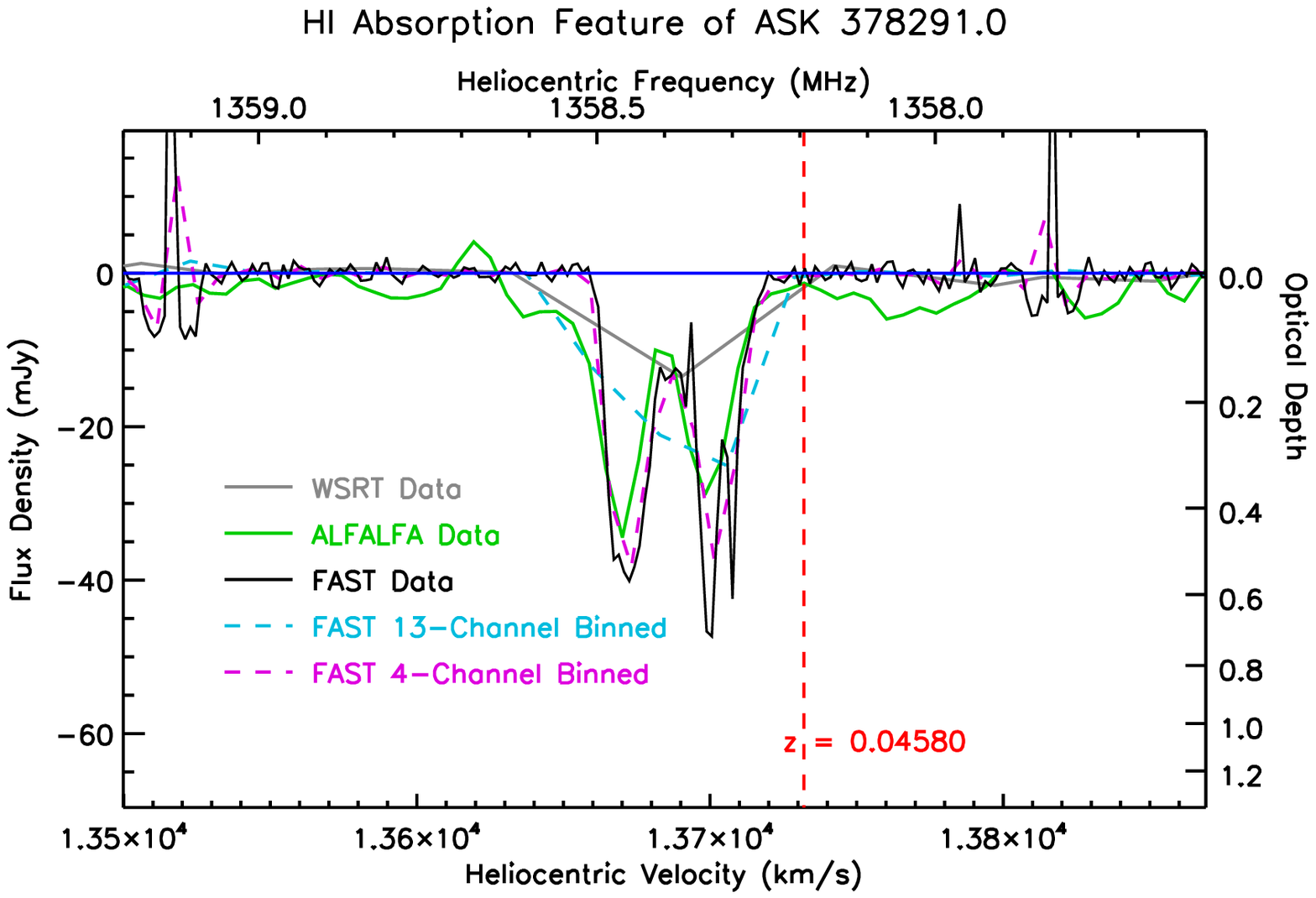}
\caption{Comparison between FAST 19-beam, WSRT and ALFALFA observations of ASK 378291.0. The black line shows data taken by FAST, while the greenish-grey line is observations from WSRT, as shown in \citealt{Maccagni2017}, and the green line the ALFALFA data. The dotted cyan line denotes binned FAST data, with a final resolution similar to WSRT's 16 km s$^{-1}$, while the dotted magenta line the 4-channel binned FAST data to achieve a spectral resolution comparable with that of ALFALFA. The red line denotes the optical redshift, and the blue line marks the zero level. }
\label{figASK_WSRT}
\end{figure}

The ASK 378291.0 at redshift $z= 0.04580$ has been classified as a newly-defined Fanaroff-Riley type 0 \citep{Baldi2018} galaxy by \cite{Cheng2018}, with a matched NVSS source exhibiting a $\sim 76.6$ mJy flux at the 1.4 GHz band, and a well-aligned jet and counterjet \citep{Cheng2018} on its position. Its absorption line was detected by both of the wide band receiver, as well as the 19-beam array of FAST. As shown in Fig. \ref{figASK}, the frequency and the double-horned profile in data from FAST and Arecibo show rough agreements to each other, thus proving the robustness of FAST detections. However, the low frequency peak in FAST data shows a significantly larger depth than ALFALFA's results. As can be seen in Fig. \ref{figASK_WSRT}, this discrepancy exist even if the FAST data are binned to resolution similar to that of ALFALFA. Considering the fact that spectra from both of the 19-beam and the wide band receiver of FAST unveiled similar line profiles, such a discrepancy between data acquired by the two telescopes could be due to footprint positions in the ALFALFA survey, since the source might not pass through the right center of the receiver beam during those drift scans, thus leading to off-axis measurements, as well as compromised flux and line profile estimations. Besides, the same peak in the 19-beam data shows a clear spike at $~1358.3$ MHz, which cannot be identified in the spectrum taken by the wide band receiver. Since multiple RFI features exist within the desired frequency range, it is reasonable to assume that such a spike is due to undesired interference overlaid on the double-horned absorption line. Thus, this spike was excluded in the fitting procedure of the line shape, as well as optical depth calculation.  

Compared with the results listed in Song et al. (2021, in preparation), which adopted a single-Gaussian function for line profile fitting, and yielding a line depth of $S_{HI,peak} \sim -23.5$ mJy, a full width at half maximum (FWHM) value of $\sim 56.3$ km s$^{-1}$, and $\int \tau {\rm d} v \approx 21.72$ km s$^{-1}$, the FAST observations of ASK 378291.0 show a slightly smaller integrated optical depth, along with a narrower FWHM. However, it should be noted that the ALFALFA fitting result is based on the continuum flux provided by the VLA Faint Images of the Sky at Twenty-Centimeters (FIRST) survey \citep{Becker1995}, which is $73.85$ mJy. When adopting this lower flux level, the FAST observations can lead to a slightly larger line depth than ALFALFA, which is $\int \tau {\rm d} v \approx 25.53 \pm 0.94$ km s$^{-1}$.

It should also be noted that this absorption feature has also been detected by the WSRT absorption line survey, with an FWHM of $\sim 66.95$ km s$^{-1}$, and  $\int \tau {\rm d} v \approx 9.62$ km s$^{-1}$ \cite{Maccagni2017}. Fig. \ref{figASK_WSRT} show a comparison between the FAST and WSRT data. It can be seen that the WSRT data only exhibit a single-peaked structure, rather than a double-horned one as shown by FAST and ALFALFA. And the maximum absorption measured by WSRT is only $-13.55$ mJy, which is nearly 3.4 times less than measurements performed by FAST. When channel-binned to a velocity resolution equivalent to WSRT, a single-peaked absorption line emerges in the FAST data, although with a lopsided profile, and a deeper feature of $-25.08$ mJy. Thus, even adopting the higher continuum flux value from \cite{Maccagni2017} ($S_{1400 \text{ MHz}} \sim 92.83$ mJy) compared with NVSS or VLA FIRST catalogues, the FAST data can still lead to an $\int \tau {\rm d} v \approx 17.22$ km s$^{-1}$, which is nearly 1.8 times higher than the WSRT integrated optical depth. Such a clear difference could be attributed to the incomplete $u-v$ coverage of a radio telescope array, which may result in compromised flux measurements. 

\subsubsection{UGC 00613}

\begin{figure}
\includegraphics[width=\columnwidth]{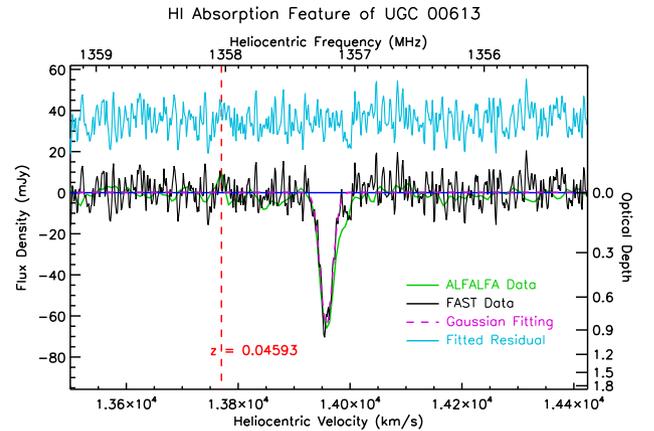}
\caption{HI absorption feature of UGC00613 obtained by the FAST wide band receiver. The black line shows observations taken by the FAST, while the grey curve is data from the ALFALFA survey. The magenta dotted line is the fitting result with a single Gaussian function, while the cyan line shows the fitting residual with a 35 mJy offset. It can be seen that the structure of this absorption feature observed by the two telescopes show good agreements with each other, although Arecibo data show a slightly broader wing component in the lower frequency end.} 
\label{figUGC}
\end{figure}

The absorption feature in UGC 00613, a flat spectrum radio galaxy with extremely diffuse envelope and faint lobes at $z = 0.04593$ \citep{Condon1991}, was first reported by \cite{Wu2015} and Song et al. (2021, in preparation), with a line depth of $S_{HI,peak} \sim -65.0$ mJy, an FWHM of $\sim 33.6$ km s$^{-1}$, and $\int \tau {\rm d} v \approx 31.14$ km s$^{-1}$. The FAST telescope performed tracking observations on this source with during its early-science phase, with its wide band receiver. Due to the lack of reliable noise diode in this period, the flux of UGC 00613 obtained by the N9020A MXA spectrum analyser were estimated in reference to observations of ASK 378291.0 shown in the upper panel of Fig. \ref{figASK}. Firstly, the baseline-subtracted wide-band-receiver data of ASK 378291.0 (lower panel of Fig. \ref{figASK}) was re-scaled according to the flux level of the same source measured by the 19-beam receiver; then, the same scaling factor for conversion from N9020A MXA's instrumental reading to flux density (measured in mJy) was applied to UGC 00613 observations. Although such a calibration process is not strictly accurate, the resulted line depth, $\sim -64.34$ mJy, is in good agreements with the Arecibo data. The discrepancy between line width estimations of the two telescopes may arise from the higher spectral resolution of FAST, which could lead to a better estimation of the line profile; as well as the relatively high RMS level in the N9020A MXA data ($\sim 6.49$ mJy, which is an order-of-magnitude higher than that of the 19-beam data), which brings more uncertainties in identification of the wing structure, as can be seen from Fig. \ref{figUGC}.

It should be noted that as can be seen in Fig. \ref{figUGC}, and also noted by Song et al. (2021, in preparation), the centre of the UGC 00613 absorption feature is redshifted from the optical redshift by as many as $\sim 186$ km s$^{-1}$ ($\sim 188$ km s$^{-1}$ for Arecibo data). According to \cite{Wegner1999}, although the signal-to-noise ratio of the optical data used for $z$ estimation is not high enough to accurately measure the velocity dispersion of the spectral lines, the redshift of this galaxy can be reliably determined as $ cz ~ 13770 \pm 23$ km s$^{-1}$. Thus, it is unlikely that such a large offset (which is at $>8 \sigma$ level of the optical measurement) between HI line and optical velocities arise from observation errors. Rather, this discrepancy may indicate the existence of unsettled infalling gas cloud or other similar structures around this AGN (e.g., see \citealt{Maccagni2017} and references herein), and further interferometric spectral line observation is required to reveal the details of the absorbing gas in this galaxy.

\subsubsection{CGCG 049-033}

The $z=0.04464$ elliptical galaxy CGCG 049-033 in Abell 2040 cluster possesses a highly asymmetric Fanaroff-Riley type II AGN, comprising of one of the largest radio jet yet discovered, a $> 10^9 M_{{\rm Sun}}$ central black hole, and intense polarised synchrotron radiation extending to a distance of $\sim 440$ kpc away from the galactic centre \citep{Bagchi2007}. The HI absorption feature with a maximum line depth $S_{HI,peak} \sim -4.3$ mJy, an FWHM $\sim 123.9$ km s$^{-1}$, and $\int \tau {\rm d} v \approx 7.23$ km s$^{-1}$ reported by Song et al. (2021, in preparation) is consistent with the general trend that early-type galaxies are in lack of abundant gaseous contents.

\begin{figure}
\includegraphics[width=\columnwidth]{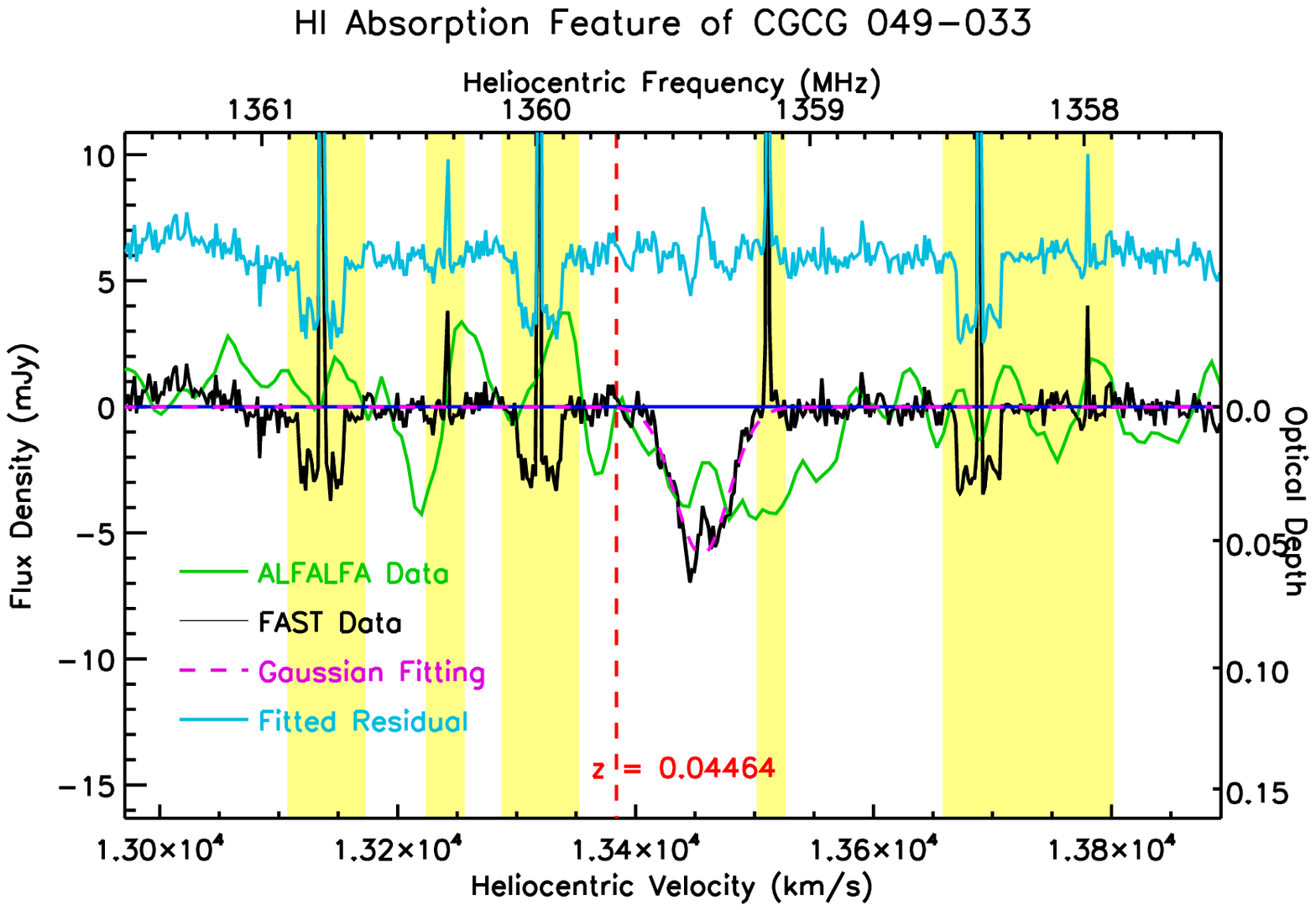}
\caption{HI absorption feature of CGCG 049-033 obtained by the FAST 19-beam receiver on October 31, 2018. Legends are the same as the upper panel of  Fig. \ref{figASK}, except that the line profile is fitted with a single Gaussian function, and the offset for fitting residual is 6 mJy.} 
\label{figCGCG}
\end{figure}

The FAST telescope conducted targeted observations twice towards CGCG 049-033, on October 31st and December 28th, 2018, respectively, each was comprised of 1200 s of on source exposure time. Although weak, the expected absorption feature did appear at the right frequency channels in both set of data, thus confirming the correctness of our detection. However, since the December observations suffered stronger RFI contamination, they have not been invoked in our analysis to get the line profile, flux, optical depth, as well as other characteristics of this HI absorber. 

As it can bee seen in Fig. \ref{figCGCG}, the HI absorption line from CGCG 049-033 is the weakest among all five sources detected by Arecibo and confirmed by FAST, with a maximum depth of $5.851$ mJy only, as measured by FAST. However, the integrated optical depth, $3.069$ km s$^{-1}$, is less than half of that of the ALFALFA results. Even if the lower continuum flux provided by VLA FIRST, as adopted by Song et al. (2021, in preparation), is applied to FAST data, the resulting $\int \tau {\rm d} \nu$ is still smaller than ALFALFA. This discrepancy should be the result of the much narrower line structure observed by FAST, and the spike-like RFI right on the low-frequency part of the absorption line further affects the fitting of the line profile, resulting in a FWHM less than half of Arecibo's results. Since the signal-to-noise ratio for this source in ALFALFA data is quite low, considering the RMS level $\sim 2.2$ mJy is almost the half of the peak estimation by Song et al. (2021, in preparation), the ALFALFA line profile itself exhibit considerable uncertainties.  Thus, the low-noise FAST data should be considered more reliable, thanks to the better sensitivity and spectral resolution of the FAST telescope.

\subsubsection{J1534+2513}

The existence of HI absorption feature in radio source J1534+2513 at a redshift of $z=0.03396$ was first proposed by \cite{Haynes2011}, and later confirmed by \cite{Wu2015} and Song et al. (2021, in preparation), with a line depth of $S_{HI,peak} \sim -23.0$ mJy, FWHM $\sim 26.4$ km s$^{-1}$, and $\int \tau {\rm d} v \approx 18.31$ km s$^{-1}$. This absorber has also been detected by the WSRT survey, with a wider line width (FWHM $\sim 40.88$ km s$^{-1}$), and an integrated optical depth of $10.92$ km s$^{-1}$, against a $43.39$ mJy background \citep{Maccagni2017}.

\begin{figure}
\includegraphics[width=\columnwidth]{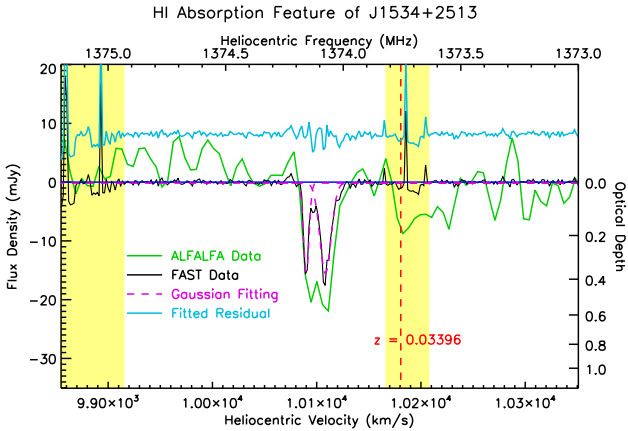}
\caption{HI absorption feature of J1534+2513 obtained by the FAST 19-beam receiver. Legends are the same as the upper panel of Fig. \ref{figASK}, with an offset for the fitting residual set as 8 mJy.} 
\label{figJ1534}
\end{figure}

\begin{figure}
\includegraphics[width=\columnwidth]{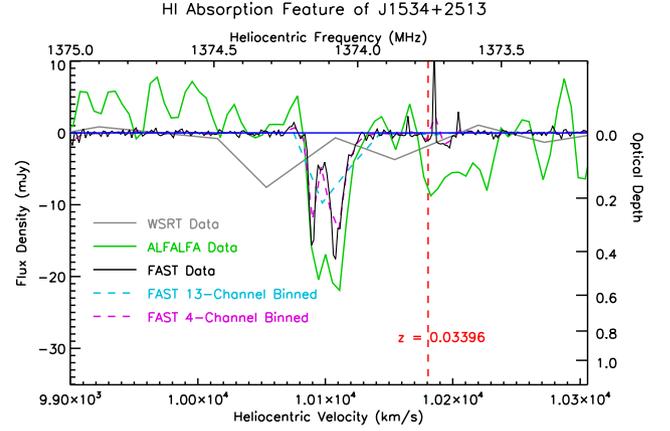}
\caption{Comparison between FAST 19-beam, WSRT and ALFALFA observations of J1534+2513. The legends are the same as Fig \ref{figASK_WSRT}. The red line denotes the optical redshift, while the blue line marks the zero level.}
\label{figJ1534_WSRT}
\end{figure}

As can be seen in Fig. \ref{figJ1534}, the most prominent feature clearly shown in the FAST spectrum should be the double-horned structure over a relatively narrow $\sim 27.53$ km s$^{-1}$ frequency range, which is only barely seen with the $\sim5.3$ km s$^{-1}$ resolution of the ALFALFA data (Song et al. 2021, in preparation), and the $\sim 16$ km s$^{-1}$-resolution WSRT survey (\cite{Maccagni2017}, see Fig. \ref{figJ1534_WSRT}), thus demonstrating the necessity of performing observations with finer spectral resolution. And the centroid of this absorption line is again significantly deviates from the optical redshift, although in this case, the radio line is blueshifted by ~ $82.15$ km s$^{-1}$ compared with the optical $z$, although the extent of discrepancy is smaller than that of UGC 00613. Considering the error in $cz$ measurement for J1534+2513 is as small as $<2.7$ km s$^{-1}$ (which is the smallest among all five sources discussed in this work), such a line velocity difference must be originated from the absorbing HI content itself. Again, this may indicate the existence of HI outflows along the line-of-sight, and inteferometer observations should be of help to unveil the nature of the absorbing gas. 

Compared with Arecibo and WSRT data, the spectral parameters of J1534+2513 obtained with FAST exhibit a narrower width, a shallower peak depth, as well as a smaller optical depth, compared with the other 2 sets of data. And although the FAST line profile is significantly deeper than WSRT's $<10$ mJy (which still holds true if these data are binned to WSRT resolution), the line depth taken by ALFALFA survey is even larger. Besides, the distance between the two spectral peaks in the WSRT data is much larger than FAST and Arecibo's results. In one sentence, the existing 3 data sets do not coincide with each other, even when taken spectral resolution into considerations. A possible explanation for such an inconsistency between different instruments could be the errors induced by a relatively high RMS level compared with line depth for ALFALFA (which has a baseline fluctuation as large as nearly 10 mJy for frequencies around the J1534+2513 absorption feature, as illustrated in Figs. \ref{figJ1534} and \ref{figJ1534_WSRT}) and WSRT, which can make their results less accurate. Also, the interferometric nature of WSRT observations could bring extra compromise to flux estimations.

\subsubsection{PGC 070403}\label{sec:PGC} 

It was \cite{Haynes2011} who put forward the first indication on the existence of $z= 0.0251$ absorption feature PGC 070403, and later the same structure has been tentatively identified by \cite{Wu2015} and Song et al. (2021, in preparation) from the $\alpha.40$ data, with an estimation of line depth as $S_{HI,peak} \sim -12.9$ mJy, FWHM $\sim 88.9$ km s$^{-1}$, and $\int \tau {\rm d} v \approx 11.11$ km s$^{-1}$. As noted by Song et al. (2021, in preparation), similar to the case of UGC 00613, this absorption line is also redshifted from the optical $cz$ by as many as $\sim 191.5$ km s$^{-1}$. However, it is also noticed that the error of optical $cz$ measurement is as large as $118.4$ km s$^{-1}$, which means that such a large discrepancy may be due to observational uncertainties.

\begin{figure}
\includegraphics[width=\columnwidth]{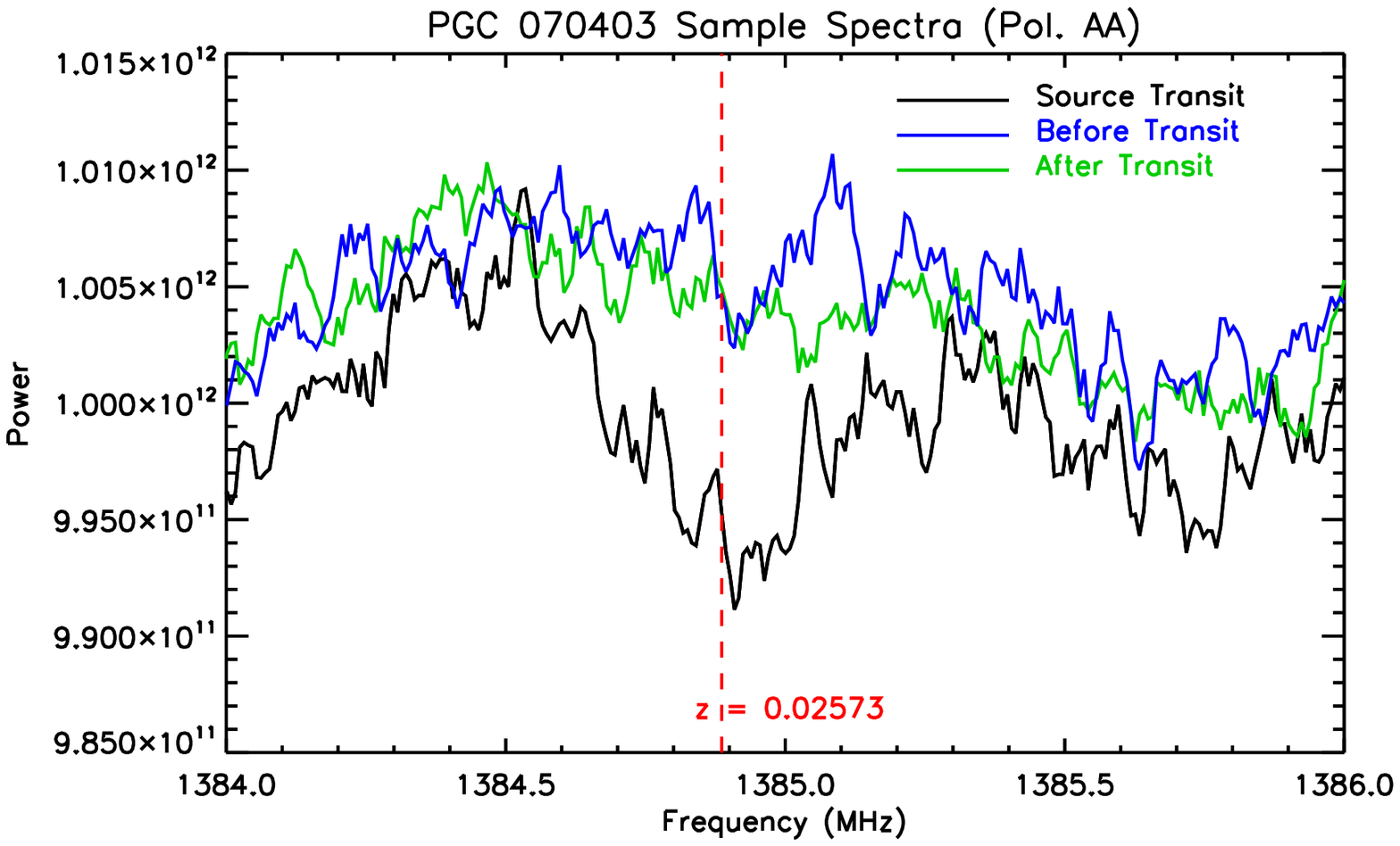}\\
\includegraphics[width=\columnwidth]{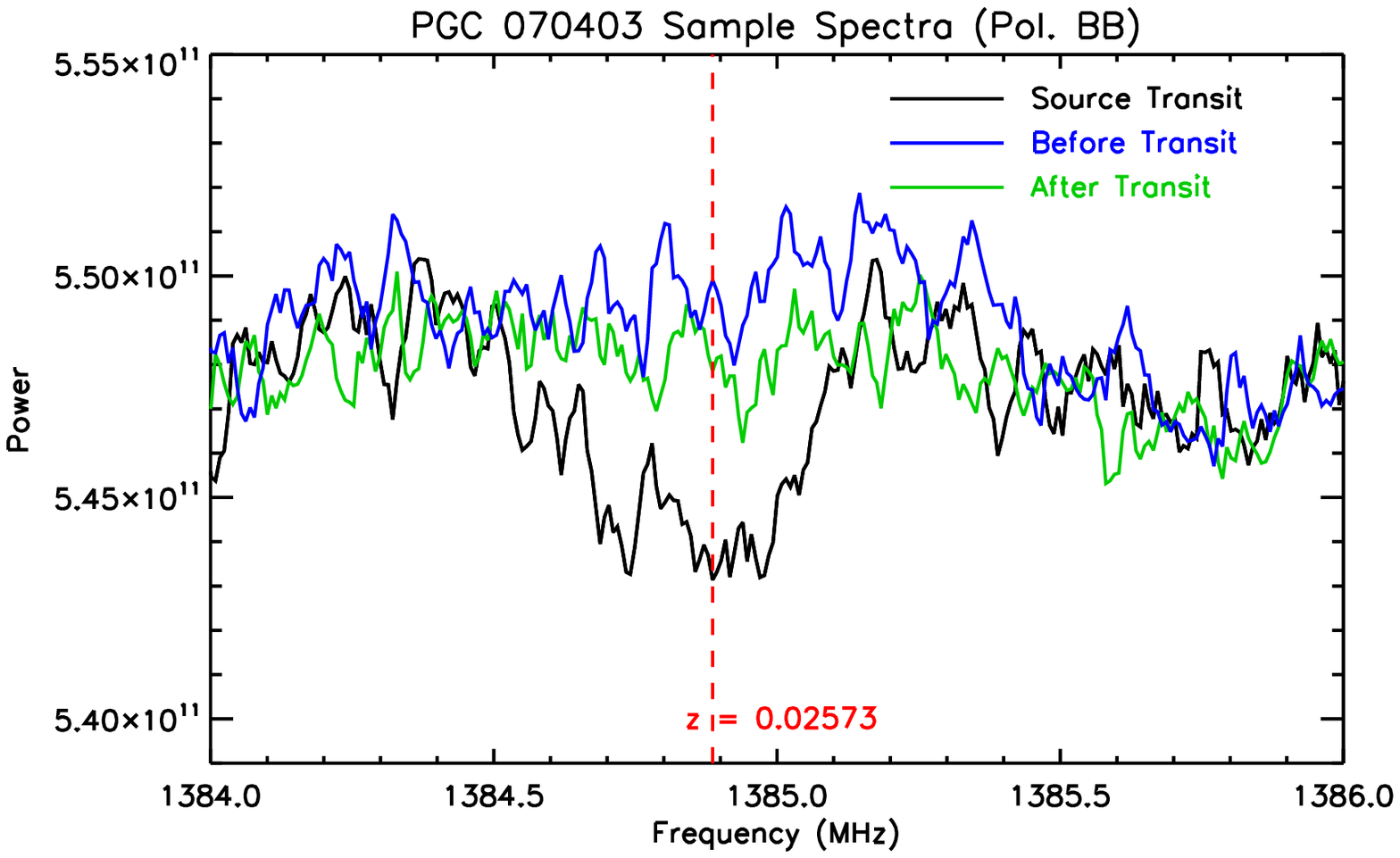}
\caption{Drift scan data of PGC 070403 from both polarisations. The y-axis is shown in the original instrumental reading. The black line shows the average of $\sim 12$ s original data taken around the transit time of the targeted source, Gaussian-smoothed by 4 spectral channels to highlight possible line features. The blue and green lines are 4-channel Gaussian-smoothed, $\sim 12$ s-averaged data taken before and after source transit, respectively, with $\sim +  10$\% of adjustments on the y-axis readings for clarity. It can be seen that a ``dip'' appears at $\sim 1384.9$ MHz around the transit time on both polarisations. And the red dotted line denotes the $cz_{peak}$ of the expected absorption feature as measured by Arecibo, which is almost coincident with the position of the transit ``dip''.} 
\label{figPGC_original} 
\end{figure}

Among all of the 5 HI absorbing sources observed by the FAST telescope and mentioned in this paper, PGC 070403 is the one with the second-weakest line depth, and the only sample observed with the drift scan mode by the FAST telescope. With a beam width of $\sim 2'.9$ at $\sim 1400$ MHz band, the central feed of the 19-beam receiver can observe each source for $\sim 12$ s during a one-pass scan. As can be seen in Fig. \ref{figPGC_original}, the absorption ``dip'' only appear during the $\sim 12$ s period around the transit time of PGC 070403 in both polarisations, thus making the FAST detection reliable. Considering the spectral baseline fluctuations appearing during the entire drift scan session, only $\sim 12$ s observations around target transit time are utilised in our data reduction process. The final spectrum is an average of the $\sim 12$ s transit data, and the background continuum is subtracted with similar smoothing and fitting method as the tracking data for sources described above.

\begin{figure}
\includegraphics[width=\columnwidth]{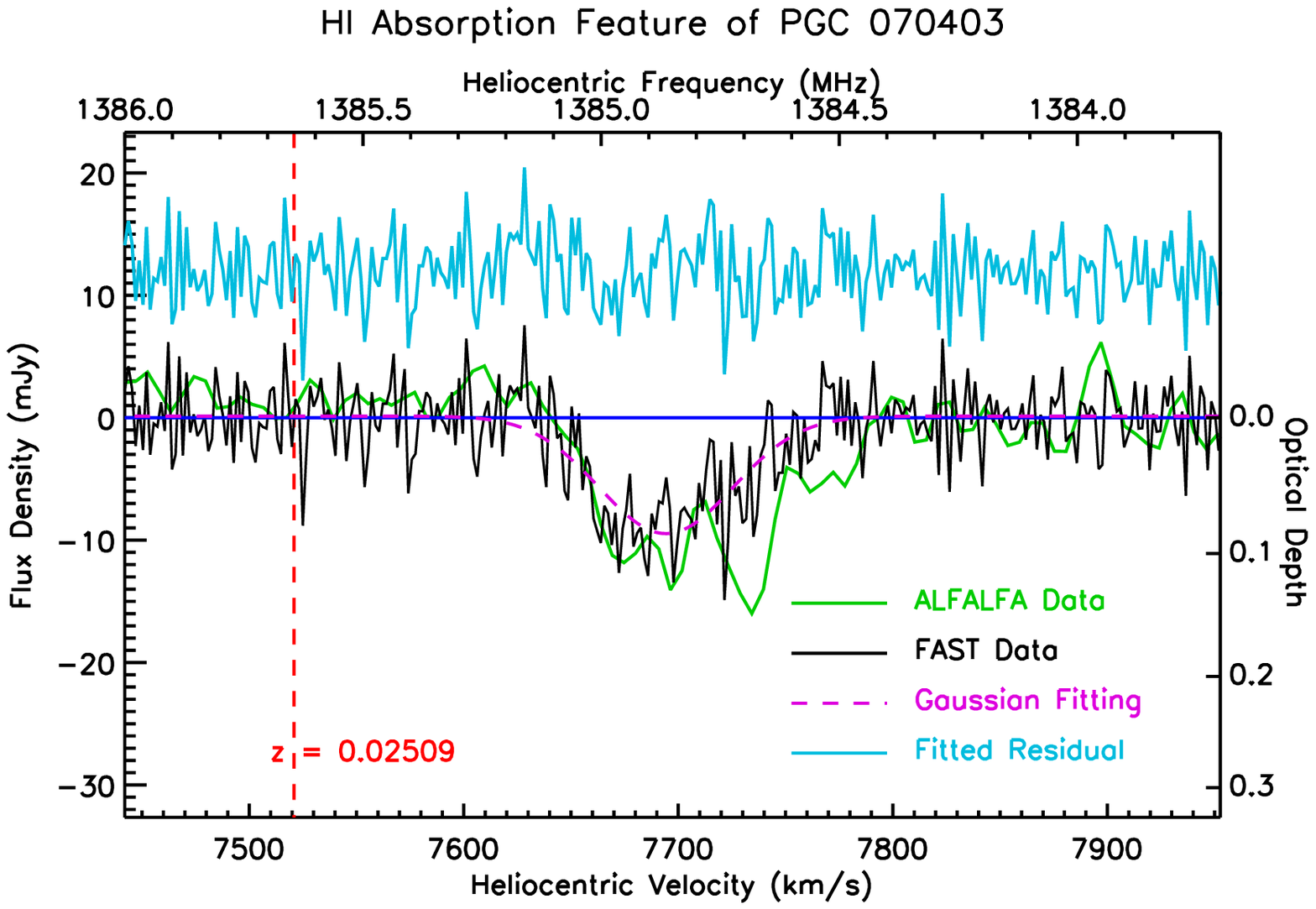}\\
\includegraphics[width=\columnwidth]{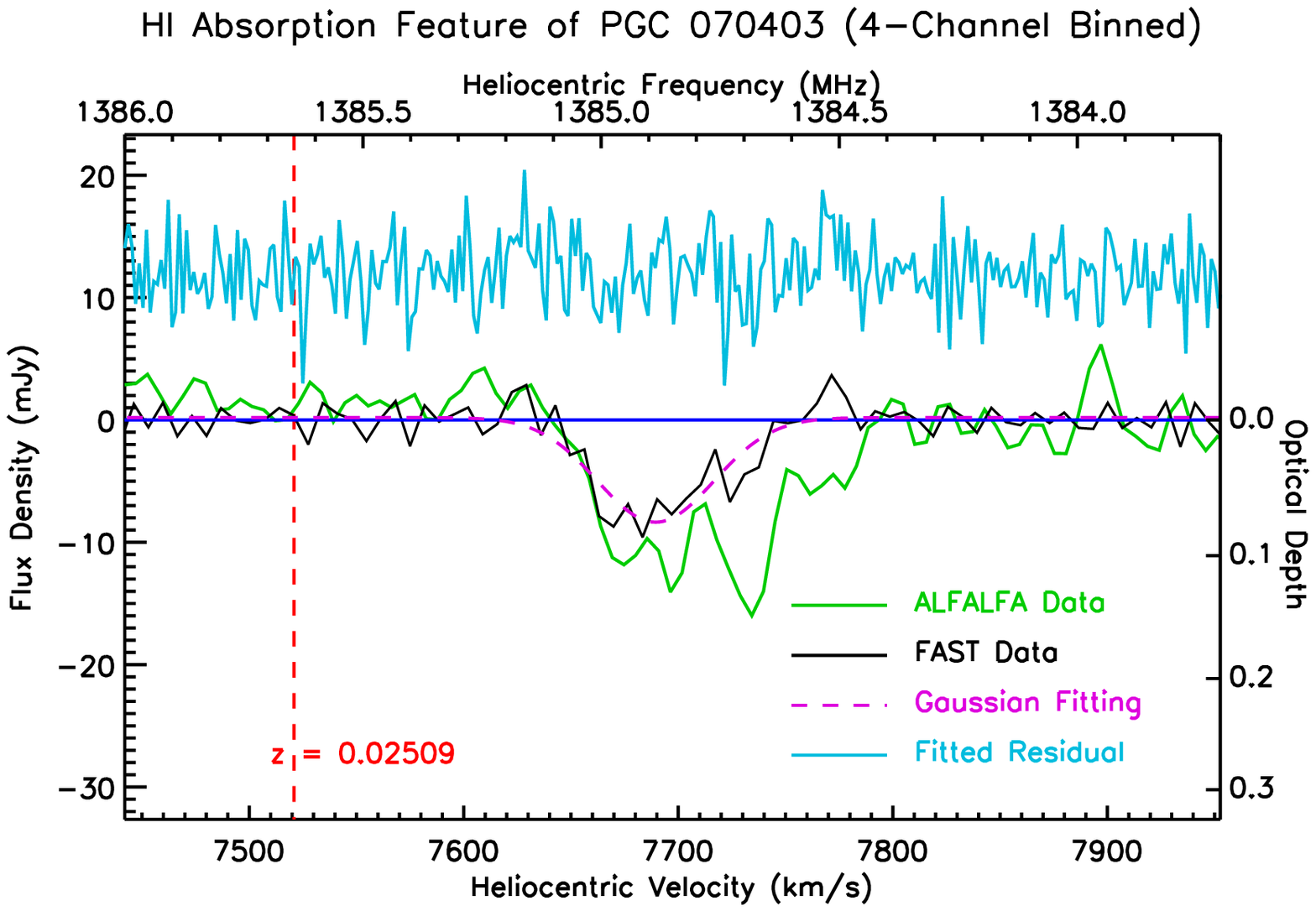}
\caption{Top: HI absorption feature PGC 070403, as observed by the 19-beam receiver of FAST. $\sim 12$ s of drift scan data around the transit time have been added up to obtain the line profile. Legends are the same as the upper panel of Fig. \ref{figCGCG}. The offset for the fitting residual is chosen as 12 mJy. Bottom: The same data, binned to a spectral resolution of $\sim 6$ km s$^{-1}$.} 
\label{figPGC}
\end{figure}

The line centre of PGC 070403 determined by FAST, which is $\sim 7694$ km s$^{-1}$, is slightly redshifted compared with ALFALFA's $7712$ km s$^{-1}$. Also, our line width is $\sim16$\% narrower than the Arecibo result, and the peak depth of the PGC 070403 absorbing feature obtained by FAST is more than $25$\% shallower than ALFALFA's value. Besides, the deepest structure in Arecibo data corresponding to a velocity of $7740$ km s$^{-1}$ does not show up in FAST observations. Combined with all of the factors mentioned above, the integrated optical depth obtained with FAST became $\sim 4.47$ km s$^{-1}$ smaller than ALFALFA's 11.11 km s$^{-1}$. 

And as shown in the bottom panel of Fig. \ref{figPGC}, with a 4-channel bin applied to the original FAST data (which leads to a spectral resolution of $\sim 6$ km s$^{-1}$, comparing to ALFALFA's $\sim 5.3$ km s$^{-1}$), an absorbing feature with line centre at $7689.07 \pm 2.14$  km s$^{-1}$, a narrower FWHM $\approx 64.37 \pm 4.29$ km s$^{-1}$, a shallower $S_{HI,peak} \sim -8.525 \pm 1.217$ mJy, and a smaller $\int \tau {\rm d} v \approx 4.783 \pm 0.809$ km s$^{-1}$, corresponding to an HI column density $N_{HI}\sim \left(8.719 \pm  1.570\right)\times 10^{20}\left(T_s/100\text{ K}\right)$ cm$^{-2}$, imposed on a background RMS level of $\sim 1.21$ mJy. Such a set of parameters exhibits noticeable differences with the ones acquired with the original, unbinned data. 

As for the discrepancy of the low-frequency part of the line profiles between FAST and Arecibo observations, we suggest that polarisation-dependent weak RFI or baseline fluctuations, which is quite common for FAST data (e.g., see Fig. 26 in \citealt{Jiang2020}), could be a possible explanation. As can be seen in Fig. \ref{figPGC_original}, compared with polarisation A, the uncalibrated spectrum in polarisation B shows a broader profile, as well as a greater depth in the low frequency regime, and is more similar to the ALFALFA result. When polarisation-combined, the $\sim 7740$ km s$^{-1}$ dip is largely missed due to polarisation-dependent behaviours. Such an effect influences more for weak sources, especially when the integration time is limited. Thus, due to complications induced by high background noise level of drift scan observations, targeted long exposures with FAST or other sensitive radio telescopes should be needed to pin down the characteristics for PGC 070403. Anyway, current results still demonstrate the feasibility of identifying weak HI absorption sources via FAST drift scan observations, thus more similar detections can be expected with the upcoming CRAFTS survey.

\section{Discussions}\label{sec:4}

\subsection{Note on ALFALFA detection rate}

In our pilot study of extragalactic HI absorbers with the FAST telescope during its commissioning phase, all of the 5 sources firstly discovered with 40\% of the ALFALFA data have been confirmed. Combining the 5 known sources identified by \cite{Wu2015} and Song et al. (2021, in preparation), the detection rate for HI absorption features in $\alpha.40$, which is $\sim 5.5$\%, still holds, assuming the radio luminosity function of radio-loud AGNs to be in the form of that from \cite{Mauch2007}. Such a rate is higher than the result estimated by \cite{Darling2011} with 7.4\% of ALFALFA data, which is $\sim 3$\%, as well as a later prediction by \cite{Allison2014} ($\sim 1.6-4.4$)\%.

However, as noted in \cite{Saintonge2007} and \cite{Haynes2011}, the performance of the matched-filtering algorithm adopted by the source finding process drops significantly below an S/N value of $\sim 6.5$, which means that weak sources embedded in data cannot be identified completely, while \cite{Wu2015} detected all the 10 absorbers based on the same method described by \cite{Saintonge2007}. Since sources CGCG 049-033, PGC 070403, as well as 3 other known sources identified with $\alpha.40$ exhibit flux levels less than the signal-to-noise ratio threshold, the detection rate of 5.5\% can only be considered as a lower limit when estimating the total number of extragalactic HI absorbers. However, since the rate for efficient detection (${\rm S}/{\rm N} > 6.5$) is 2.75\%, which is only the half of the 5.5\% value, it is safe to say that at least 12 or 13 HI absorbing systems can be identified with the $\alpha.100$ data (\citealt{Haynes2018}), and the total number of detections is hard to predict due to algorithm limitations.

\subsection{Implications on the prospect of HI absorption line detections with FAST}
Based on the different methods we have tested during the pilot observations, we find that a single-pass scan can already resolve weak absorption, while the 2-pass drift scan mode similar to that of ALFALFA \citep{Giovanelli2005} would be a more efficient strategy to detect large number of HI absorbers, since in this case, more time-varied RFI and other fluctuations can be excluded. In fact, the 2-pass strategy is the observing mode already adopted by the FAST extragalactic HI survey, which is one of the key projects undertaken at FAST, thus we can make good use of its large amount of data sets in the near future. Since it takes a single source $\sim 12$ s to pass through one beam during each scan, and a rotation angle of $23^{\circ}.4$ would be applied to the 19-beam receiver \citep{Li2018} to ensure maximum Declination coverage and non-overlap between all feed horns during sky surveys, the total integration time for each source with all scans finished should be $\sim 24$ s. We calculate the mean RMS value for a 24 s integration duration with all of the 19-beam data analysed by this paper, leading to a averaged noise level of $\sim 3$ mJy, which is comparable to the sensitivity of FLASH survey ($\sim 3.2$ mJy beam$^{-1}$) achieved with 2 hr of integration time by ASKAP (\citealt{Allison2020}). And such a noise level can be further suppressed to $\sim 1.5$ mJy with an extra 4-channel bin along the frequency (velocity) direction, which is slightly lower than the noise level of WALLABY ($\sim 1.6$ mJy beam$^{-1}$ with $8\times 2$ hr ASKAP integration time, see \citealt{Koribalski2020}). Thus, in the following evaluations, we calculate the prospect of HI absorption detections for FAST sky surveys with 2 sensitivity levels, $\sim 3$ and $\sim 1.5$ mJy, respectively.  
Take the averaged optical depth for 10 HI absorbers (including 5 previously-known samples) presented in Song et al. (2021, in preparation), $\tau \sim 0.352$, as a typical value for extragalactic HI absorption features, the corresponding normalised line depth should be of $\sim 0.3$ times the continuum flux. And suppose a reliable identification requires a $\geqslant 5 \sigma$ detection, averagely speaking, a continuum source, such as AGN, which served as the background for HI absorption, is required to have a minimum flux of $\sim 38$ mJy for detecting with the original unsmoothed observations, and $\sim 25$ mJy for the 4-channel binned data.

The maximum zenith angle that can be attained by FAST is $\sim 40^{\circ}$ (\citealt{Nan2011, Li2018, Jiang2020}). Given the telescope's geographic latitude at $25^{\circ} 39' 10''.6$ N, the observable Declination extent of FAST should be $\sim -14^{\circ}.35<\delta<65^{\circ}.65$. Thus, the full FAST observable sky covers an area of $\sim 23800 $ deg$^2$, which is more than 3 times the mapped region of the ALFALFA survey (\citealt{Giovanelli2005, Haynes2018}). And the 19-beam receiver, the primary workhorse for sky surveys, operates at an HI redshift range of $\sim -6000 < cz < 105000$ km s$^{-1}$, compared to ALFALFA's $-2000$ to $18000$ km s$^{-1}$ \citep{Haynes2018}. Combined with both factors, the observable comoving volume of FAST should be $\sim4.2$ Gpc$^3$, which is nearly $\sim 300$ times the ALFALFA's total coverage.  

Following similar procedures by previous works such as \cite{Allison2014} and \cite{Wu2015}, we adopt the local luminosity function for radio-loud AGNs proposed by \cite{Mauch2007}. Ignoring the dependence of AGN distributions on redshift, we predict a total number of $\sim 43,000$ AGNs with fluxes above the $5 \sigma$ limit in the complete FAST sky survey coverage with a $\sim 1.5$ km s${^-1}$ spectral resolution, and $\sim 49,000$ AGNs for the 4-channel smoothed data. Applying the detection rate of HI absorbing systems calculated by Song et al. (2021, in preparation), $\sim 2,300$ extragalactic HI absorption lines should be identified with the complete set of original FAST sky survey data, and $\sim 2,600$ such sources could be detected with the 4-channel binned spectra. Of course, the calculations above only provide the most optimistic expectations, and the more or less compromised antenna gain of FAST for zenith angles larger than $\sim 26^{\circ}.4$ (\citealt{Li2018, Jiang2020}) was not taken into account. Besides, a number of frequencies in the $1050-1450$ MHz band are often heavily contaminated by RFI generated by distance measuring equipments for aviation, or emitters on board navigation satellites. Thus, the real number of HI absorption line detection could be largely reduced. However, even neglecting the frequency band of $\sim 1150-1300$ MHz, which is most severely affected by RFI from satellites, $\sim 1,500$ HI absorption systems can still be expected from the unbinned data, and $\sim 1,700$ from the 4-channel binned data, which is consistent with the forecast made by \cite{Yu2017}.

Another approach to predict the FAST detection prospect for HI absorbers is with the NVSS (\citealt{Condon1998}) source count. $\sim 1.2 \times 10^5$ continuum sources with $S_{1.4 \text{ GHz}}>38$ mJy exist within the FAST observable sky, and the number for sources with $S_{1.4 \text{ GHz}}>25$ mJy is $\sim 1.8 \times 10^5$. With a detection rate of $\sim5.5$\% applied, these numbers can lead to $\sim 6000$ extragalactic HI absorbers found within the original data, and $\sim 9000$ such systems for 4-channel binned spectra. Still, since the NVSS catalogue does not provide associated redshift information for each individual source, and the chance of alignment between a high-$z$ continuum source along with a low-$z$, HI-bearing galaxy is low, it is almost certain that a significant amount of HI absorbers associated with NVSS sources lie beyond the frequency coverage of FAST. Thus, such an estimation can only be considered as the rough upper limit for the CRAFTS survey.

\section{Conclusions}\label{sec:5}

In this paper, we report FAST observations of 5 extragalactic HI absorption sources firstly identified from the ALFALFA survey data. We confirm the existence of all absorption features from these sources. However, the line widths, optical depths, as well as HI column densities of the detected HI absorbers as derived from FAST data show noticeable discrepancies with the results previously obtained by Arecibo and WSRT, due to various factors. Since the FAST data have much higher S/N and finer spectral resolution compared with existing sky surveys, more features of the HI absorption lines, such as the double-horned structure of J1534+2513, can be revealed with high confidence. And the HI absorption line of PGC 070403, which exhibits the second-shallowest line depth among the 5, was successfully detected during a $\sim 12$ s integration time using the drift scan mode, which will be the final choice for the upcoming extragalactic HI surveys. These observations, which can be considered as the first batch of extragalactic absorption line detections performed by FAST, demonstrated the capability of this telescope for HI absorption studies. It is expected that with a larger sky coverage and higher sensitivity than Arecibo, over 1,500 extragalactic HI absorbers could be unveiled with the entire set of FAST extragalactic sky survey data.

\section*{Acknowledgements}

This work is supported by the National Key R\&D Program of China (Grant No. 2017YFA0402600), the National Natural Science Foundation of China (Grant Nos. 11903056, 11763002), the Joint Research Fund in Astronomy (Grant Nos. U1731125, U1531246, U1931203) under cooperative agreement between the National Natural Science Foundation of China (NSFC) and Chinese Academy of Sciences (CAS), the Cultivation Project for FAST Scientific Payoff and Research Achievement of CAMS-CAS, as well as the Open Project Program of the Key Laboratory of FAST, National Astronomical Observatories, Chinese Academy of Sciences (NAOC). The authors thank Prof. Martha P. Haynes for providing the ALFALFA HI absorption spectra, and Cheng Cheng, You-Gang Wang, Hong-Liang Yan, as well as Si-Cheng Yu for helpful discussions. Also, the authors would like to thank the anonymous referee for valuable comments.  

\section*{Data Availability}

The data analysed by this work are from the FAST project Nos. 3017 (UGC 00613, ASK 378291.0, CGCG 049-033, J1534+2513) and N2020\_3 (PGC 070403), and can be accessed by sending request to the FAST Data Centre or to the corresponding authors of this paper.



\bibliographystyle{mnras}
\bibliography{example} 

\begin{thebibliography}{}
\makeatletter
\relax
\def\mn@urlcharsother{\let\do\@makeother \do\$\do\&\do\#\do\^\do\_\do\%\do\~}
\def\mn@doi{\begingroup\mn@urlcharsother \@ifnextchar [ {\mn@doi@}
  {\mn@doi@[]}}
\def\mn@doi@[#1]#2{\def\@tempa{#1}\ifx\@tempa\@empty \href
  {http://dx.doi.org/#2} {doi:#2}\else \href {http://dx.doi.org/#2} {#1}\fi
  \endgroup}
\def\mn@eprint#1#2{\mn@eprint@#1:#2::\@nil}
\def\mn@eprint@arXiv#1{\href {http://arxiv.org/abs/#1} {{\tt arXiv:#1}}}
\def\mn@eprint@dblp#1{\href {http://dblp.uni-trier.de/rec/bibtex/#1.xml}
  {dblp:#1}}
\def\mn@eprint@#1:#2:#3:#4\@nil{\def\@tempa {#1}\def\@tempb {#2}\def\@tempc
  {#3}\ifx \@tempc \@empty \let \@tempc \@tempb \let \@tempb \@tempa \fi \ifx
  \@tempb \@empty \def\@tempb {arXiv}\fi \@ifundefined
  {mn@eprint@\@tempb}{\@tempb:\@tempc}{\expandafter \expandafter \csname
  mn@eprint@\@tempb\endcsname \expandafter{\@tempc}}}

\bibitem[\protect\citeauthoryear{{Adams} et~al.,}{{Adams}
  et~al.}{2018}]{Adams2018}
{Adams} E.,  et~al., 2018, in American Astronomical Society Meeting Abstracts
  \#231. p. 354.04

\bibitem[\protect\citeauthoryear{{Allison} et~al.,}{{Allison}
  et~al.}{2012}]{Allison2012}
{Allison} J.~R.,  et~al., 2012, \mn@doi [\mnras]
  {10.1111/j.1365-2966.2012.21062.x}, \href
  {https://ui.adsabs.harvard.edu/abs/2012MNRAS.423.2601A} {423, 2601}

\bibitem[\protect\citeauthoryear{{Allison}, {Sadler}  \& {Meekin}}{{Allison}
  et~al.}{2014}]{Allison2014}
{Allison} J.~R.,  {Sadler} E.~M.,   {Meekin} A.~M.,  2014, \mn@doi [\mnras]
  {10.1093/mnras/stu289}, \href
  {https://ui.adsabs.harvard.edu/abs/2014MNRAS.440..696A} {440, 696}

\bibitem[\protect\citeauthoryear{{Allison}, {Zwaan}, {Duchesne}  \&
  {Curran}}{{Allison} et~al.}{2016}]{Allison2016}
{Allison} J.~R.,  {Zwaan} M.~A.,  {Duchesne} S.~W.,   {Curran} S.~J.,  2016,
  \mn@doi [\mnras] {10.1093/mnras/stw1722}, \href
  {https://ui.adsabs.harvard.edu/abs/2016MNRAS.462.1341A} {462, 1341}

\bibitem[\protect\citeauthoryear{{Allison} et~al.,}{{Allison}
  et~al.}{2020}]{Allison2020}
{Allison} J.~R.,  et~al., 2020, \mn@doi [\mnras] {10.1093/mnras/staa949}, \href
  {https://ui.adsabs.harvard.edu/abs/2020MNRAS.494.3627A} {494, 3627}

\bibitem[\protect\citeauthoryear{{Bagchi}, {Gopal-Krishna}, {Krause}  \&
  {Joshi}}{{Bagchi} et~al.}{2007}]{Bagchi2007}
{Bagchi} J.,  {Gopal-Krishna} {Krause} M.,   {Joshi} S.,  2007, \mn@doi [\apjl]
  {10.1086/524220}, \href
  {https://ui.adsabs.harvard.edu/abs/2007ApJ...670L..85B} {670, L85}

\bibitem[\protect\citeauthoryear{{Baldi}, {Capetti}  \& {Massaro}}{{Baldi}
  et~al.}{2018}]{Baldi2018}
{Baldi} R.~D.,  {Capetti} A.,   {Massaro} F.,  2018, \mn@doi [\aap]
  {10.1051/0004-6361/201731333}, \href
  {https://ui.adsabs.harvard.edu/abs/2018A&A...609A...1B} {609, A1}

\bibitem[\protect\citeauthoryear{{Barnes} et~al.,}{{Barnes}
  et~al.}{2001}]{Barnes2001}
{Barnes} D.~G.,  et~al., 2001, \mn@doi [\mnras]
  {10.1046/j.1365-8711.2001.04102.x}, \href
  {https://ui.adsabs.harvard.edu/abs/2001MNRAS.322..486B} {322, 486}

\bibitem[\protect\citeauthoryear{{Becker}, {White}  \& {Helfand}}{{Becker}
  et~al.}{1995}]{Becker1995}
{Becker} R.~H.,  {White} R.~L.,   {Helfand} D.~J.,  1995, \mn@doi [\apj]
  {10.1086/176166}, \href
  {https://ui.adsabs.harvard.edu/abs/1995ApJ...450..559B} {450, 559}

\bibitem[\protect\citeauthoryear{{Carilli}, {Menten}, {Reid}, {Rupen}  \&
  {Yun}}{{Carilli} et~al.}{1998}]{Carilli1998}
{Carilli} C.~L.,  {Menten} K.~M.,  {Reid} M.~J.,  {Rupen} M.~P.,   {Yun} M.~S.,
   1998, \mn@doi [\apj] {10.1086/305191}, \href
  {https://ui.adsabs.harvard.edu/abs/1998ApJ...494..175C} {494, 175}

\bibitem[\protect\citeauthoryear{{Cheng} \& {An}}{{Cheng} \&
  {An}}{2018}]{Cheng2018}
{Cheng} X.~P.,  {An} T.,  2018, \mn@doi [\apj] {10.3847/1538-4357/aad22c},
  \href {https://ui.adsabs.harvard.edu/abs/2018ApJ...863..155C} {863, 155}

\bibitem[\protect\citeauthoryear{{Chengalur} \& {Kanekar}}{{Chengalur} \&
  {Kanekar}}{2000}]{Chengalur2000}
{Chengalur} J.~N.,  {Kanekar} N.,  2000, \mn@doi [\mnras]
  {10.1046/j.1365-8711.2000.03793.x}, \href
  {https://ui.adsabs.harvard.edu/abs/2000MNRAS.318..303C} {318, 303}

\bibitem[\protect\citeauthoryear{{Chowdhury}, {Kanekar}  \&
  {Chengalur}}{{Chowdhury} et~al.}{2020}]{Chowdhury2020}
{Chowdhury} A.,  {Kanekar} N.,   {Chengalur} J.~N.,  2020, \mn@doi [\apjl]
  {10.3847/2041-8213/abb13d}, \href
  {https://ui.adsabs.harvard.edu/abs/2020ApJ...900L..30C} {900, L30}

\bibitem[\protect\citeauthoryear{{Condon}, {Frayer}  \& {Broderick}}{{Condon}
  et~al.}{1991}]{Condon1991}
{Condon} J.~J.,  {Frayer} D.~T.,   {Broderick} J.~J.,  1991, \mn@doi [\aj]
  {10.1086/115692}, \href
  {https://ui.adsabs.harvard.edu/abs/1991AJ....101..362C} {101, 362}

\bibitem[\protect\citeauthoryear{{Condon}, {Cotton}, {Greisen}, {Yin},
  {Perley}, {Taylor}  \& {Broderick}}{{Condon} et~al.}{1998}]{Condon1998}
{Condon} J.~J.,  {Cotton} W.~D.,  {Greisen} E.~W.,  {Yin} Q.~F.,  {Perley}
  R.~A.,  {Taylor} G.~B.,   {Broderick} J.~J.,  1998, \mn@doi [\aj]
  {10.1086/300337}, \href
  {https://ui.adsabs.harvard.edu/abs/1998AJ....115.1693C} {115, 1693}

\bibitem[\protect\citeauthoryear{{Curran}}{{Curran}}{2019}]{Curran2019}
{Curran} S.~J.,  2019, \mn@doi [\mnras] {10.1093/mnras/stz215}, \href
  {https://ui.adsabs.harvard.edu/abs/2019MNRAS.484.3911C} {484, 3911}

\bibitem[\protect\citeauthoryear{{Curran}, {Whiting}, {Murphy}, {Webb},
  {Longmore}, {Pihlstr{\"o}m}, {Athreya}  \& {Blake}}{{Curran}
  et~al.}{2006}]{Curran2006}
{Curran} S.~J.,  {Whiting} M.~T.,  {Murphy} M.~T.,  {Webb} J.~K.,  {Longmore}
  S.~N.,  {Pihlstr{\"o}m} Y.~M.,  {Athreya} R.,   {Blake} C.,  2006, \mn@doi
  [\mnras] {10.1111/j.1365-2966.2006.10677.x}, \href
  {https://ui.adsabs.harvard.edu/abs/2006MNRAS.371..431C} {371, 431}

\bibitem[\protect\citeauthoryear{{Darling}, {Macdonald}, {Haynes}  \&
  {Giovanelli}}{{Darling} et~al.}{2011}]{Darling2011}
{Darling} J.,  {Macdonald} E.~P.,  {Haynes} M.~P.,   {Giovanelli} R.,  2011,
  \mn@doi [\apj] {10.1088/0004-637X/742/1/60}, \href
  {https://ui.adsabs.harvard.edu/abs/2011ApJ...742...60D} {742, 60}

\bibitem[\protect\citeauthoryear{{Ger{\'e}b}, {Morganti}  \&
  {Oosterloo}}{{Ger{\'e}b} et~al.}{2014}]{Gereb2014}
{Ger{\'e}b} K.,  {Morganti} R.,   {Oosterloo} T.~A.,  2014, \mn@doi [\aap]
  {10.1051/0004-6361/201423999}, \href
  {https://ui.adsabs.harvard.edu/abs/2014A&A...569A..35G} {569, A35}

\bibitem[\protect\citeauthoryear{{Ger{\'e}b}, {Maccagni}, {Morganti}  \&
  {Oosterloo}}{{Ger{\'e}b} et~al.}{2015}]{Gereb2015}
{Ger{\'e}b} K.,  {Maccagni} F.~M.,  {Morganti} R.,   {Oosterloo} T.~A.,  2015,
  \mn@doi [\aap] {10.1051/0004-6361/201424655}, \href
  {https://ui.adsabs.harvard.edu/abs/2015A&A...575A..44G} {575, A44}

\bibitem[\protect\citeauthoryear{{Giovanelli} et~al.,}{{Giovanelli}
  et~al.}{2005}]{Giovanelli2005}
{Giovanelli} R.,  et~al., 2005, \mn@doi [\aj] {10.1086/497431}, \href
  {https://ui.adsabs.harvard.edu/abs/2005AJ....130.2598G} {130, 2598}

\bibitem[\protect\citeauthoryear{{Glowacki} et~al.,}{{Glowacki}
  et~al.}{2019}]{Glowacki2019}
{Glowacki} M.,  et~al., 2019, \mn@doi [\mnras] {10.1093/mnras/stz2452}, \href
  {https://ui.adsabs.harvard.edu/abs/2019MNRAS.489.4926G} {489, 4926}

\bibitem[\protect\citeauthoryear{{Grasha}, {Darling}, {Bolatto}, {Leroy}  \&
  {Stocke}}{{Grasha} et~al.}{2019}]{Grasha2019}
{Grasha} K.,  {Darling} J.,  {Bolatto} A.,  {Leroy} A.~K.,   {Stocke} J.~T.,
  2019, \mn@doi [\apjs] {10.3847/1538-4365/ab4906}, \href
  {https://ui.adsabs.harvard.edu/abs/2019ApJS..245....3G} {245, 3}

\bibitem[\protect\citeauthoryear{{Gupta} et~al.,}{{Gupta}
  et~al.}{2016}]{Gupta2016}
{Gupta} N.,  et~al., 2016, in MeerKAT Science: On the Pathway to the SKA. p.~14
  (\mn@eprint {arXiv} {1708.07371})

\bibitem[\protect\citeauthoryear{{Gupta} et~al.,}{{Gupta}
  et~al.}{2021}]{Gupta2021}
{Gupta} N.,  et~al., 2021, \mn@doi [\apj] {10.3847/1538-4357/abcb85}, \href
  {https://ui.adsabs.harvard.edu/abs/2021ApJ...907...11G} {907, 11}

\bibitem[\protect\citeauthoryear{{Haynes} et~al.,}{{Haynes}
  et~al.}{2011}]{Haynes2011}
{Haynes} M.~P.,  et~al., 2011, \mn@doi [\aj] {10.1088/0004-6256/142/5/170},
  \href {https://ui.adsabs.harvard.edu/abs/2011AJ....142..170H} {142, 170}

\bibitem[\protect\citeauthoryear{{Haynes} et~al.,}{{Haynes}
  et~al.}{2018}]{Haynes2018}
{Haynes} M.~P.,  et~al., 2018, \mn@doi [\apj] {10.3847/1538-4357/aac956}, \href
  {https://ui.adsabs.harvard.edu/abs/2018ApJ...861...49H} {861, 49}

\bibitem[\protect\citeauthoryear{{Heiles} \& {Troland}}{{Heiles} \&
  {Troland}}{2003a}]{Heiles2003a}
{Heiles} C.,  {Troland} T.~H.,  2003a, \mn@doi [\apjs] {10.1086/367785}, \href
  {https://ui.adsabs.harvard.edu/abs/2003ApJS..145..329H} {145, 329}

\bibitem[\protect\citeauthoryear{{Heiles} \& {Troland}}{{Heiles} \&
  {Troland}}{2003b}]{Heiles2003b}
{Heiles} C.,  {Troland} T.~H.,  2003b, \mn@doi [\apj] {10.1086/367828}, \href
  {https://ui.adsabs.harvard.edu/abs/2003ApJ...586.1067H} {586, 1067}

\bibitem[\protect\citeauthoryear{{Jiang} et~al.,}{{Jiang}
  et~al.}{2019}]{Jiang2019}
{Jiang} P.,  et~al., 2019, \mn@doi [Science China Physics, Mechanics, and
  Astronomy] {10.1007/s11433-018-9376-1}, \href
  {https://ui.adsabs.harvard.edu/abs/2019SCPMA..6259502J} {62, 959502}

\bibitem[\protect\citeauthoryear{{Jiang} et~al.,}{{Jiang}
  et~al.}{2020}]{Jiang2020}
{Jiang} P.,  et~al., 2020, \mn@doi [Research in Astronomy and Astrophysics]
  {10.1088/1674-4527/20/5/64}, \href
  {https://ui.adsabs.harvard.edu/abs/2020RAA....20...64J} {20, 064}

\bibitem[\protect\citeauthoryear{{Kanekar} \& {Briggs}}{{Kanekar} \&
  {Briggs}}{2003}]{Kanekar2003}
{Kanekar} N.,  {Briggs} F.~H.,  2003, \mn@doi [\aap]
  {10.1051/0004-6361:20031676}, \href
  {https://ui.adsabs.harvard.edu/abs/2003A&A...412L..29K} {412, L29}

\bibitem[\protect\citeauthoryear{{Kanekar} \& {Briggs}}{{Kanekar} \&
  {Briggs}}{2004}]{Kanekar2004}
{Kanekar} N.,  {Briggs} F.~H.,  2004, \mn@doi [\nar]
  {10.1016/j.newar.2004.09.030}, \href
  {https://ui.adsabs.harvard.edu/abs/2004NewAR..48.1259K} {48, 1259}

\bibitem[\protect\citeauthoryear{{Kanekar}, {Chengalur}  \& {Lane}}{{Kanekar}
  et~al.}{2007}]{Kanekar2007}
{Kanekar} N.,  {Chengalur} J.~N.,   {Lane} W.~M.,  2007, \mn@doi [\mnras]
  {10.1111/j.1365-2966.2007.11430.x}, \href
  {https://ui.adsabs.harvard.edu/abs/2007MNRAS.375.1528K} {375, 1528}

\bibitem[\protect\citeauthoryear{{Kanekar}, {Prochaska}, {Ellison}  \&
  {Chengalur}}{{Kanekar} et~al.}{2009}]{Kanekar2009}
{Kanekar} N.,  {Prochaska} J.~X.,  {Ellison} S.~L.,   {Chengalur} J.~N.,  2009,
  \mn@doi [\mnras] {10.1111/j.1365-2966.2009.14661.x}, \href
  {https://ui.adsabs.harvard.edu/abs/2009MNRAS.396..385K} {396, 385}

\bibitem[\protect\citeauthoryear{{Koribalski} et~al.,}{{Koribalski}
  et~al.}{2004}]{Koribalski2004}
{Koribalski} B.~S.,  et~al., 2004, \mn@doi [\aj] {10.1086/421744}, \href
  {https://ui.adsabs.harvard.edu/abs/2004AJ....128...16K} {128, 16}

\bibitem[\protect\citeauthoryear{{Koribalski} et~al.,}{{Koribalski}
  et~al.}{2020}]{Koribalski2020}
{Koribalski} B.~S.,  et~al., 2020, \mn@doi [\apss]
  {10.1007/s10509-020-03831-4}, \href
  {https://ui.adsabs.harvard.edu/abs/2020Ap&SS.365..118K} {365, 118}

\bibitem[\protect\citeauthoryear{{Li} et~al.,}{{Li} et~al.}{2018}]{Li2018}
{Li} D.,  et~al., 2018, \mn@doi [IEEE Microwave Magazine]
  {10.1109/MMM.2018.2802178}, \href
  {https://ui.adsabs.harvard.edu/abs/2018IMMag..19..112L} {19, 112}

\bibitem[\protect\citeauthoryear{{Maccagni}, {Morganti}, {Oosterloo},
  {Ger{\'e}b}  \& {Maddox}}{{Maccagni} et~al.}{2017}]{Maccagni2017}
{Maccagni} F.~M.,  {Morganti} R.,  {Oosterloo} T.~A.,  {Ger{\'e}b} K.,
  {Maddox} N.,  2017, \mn@doi [\aap] {10.1051/0004-6361/201730563}, \href
  {https://ui.adsabs.harvard.edu/abs/2017A&A...604A..43M} {604, A43}

\bibitem[\protect\citeauthoryear{{Mauch} \& {Sadler}}{{Mauch} \&
  {Sadler}}{2007}]{Mauch2007}
{Mauch} T.,  {Sadler} E.~M.,  2007, \mn@doi [\mnras]
  {10.1111/j.1365-2966.2006.11353.x}, \href
  {https://ui.adsabs.harvard.edu/abs/2007MNRAS.375..931M} {375, 931}

\bibitem[\protect\citeauthoryear{{Morganti} \& {Oosterloo}}{{Morganti} \&
  {Oosterloo}}{2018}]{Morganti2018}
{Morganti} R.,  {Oosterloo} T.,  2018, \mn@doi [\aapr]
  {10.1007/s00159-018-0109-x}, \href
  {https://ui.adsabs.harvard.edu/abs/2018A&ARv..26....4M} {26, 4}

\bibitem[\protect\citeauthoryear{{Morganti}, {Sadler}  \& {Curran}}{{Morganti}
  et~al.}{2015}]{Morganti2015}
{Morganti} R.,  {Sadler} E.~M.,   {Curran} S.,  2015, in Advancing Astrophysics
  with the Square Kilometre Array (AASKA14). p.~134 (\mn@eprint {arXiv}
  {1501.01091})

\bibitem[\protect\citeauthoryear{{Murray} et~al.,}{{Murray}
  et~al.}{2015}]{Murray2015}
{Murray} C.~E.,  et~al., 2015, \mn@doi [\apj] {10.1088/0004-637X/804/2/89},
  \href {https://ui.adsabs.harvard.edu/abs/2015ApJ...804...89M} {804, 89}

\bibitem[\protect\citeauthoryear{{Nan}}{{Nan}}{2006}]{Nan2006}
{Nan} R.,  2006, \mn@doi [Science in China: Physics, Mechanics and Astronomy]
  {10.1007/s11433-006-0129-9}, \href
  {https://ui.adsabs.harvard.edu/abs/2006ScChG..49..129N} {49, 129}

\bibitem[\protect\citeauthoryear{{Nan} et~al.,}{{Nan} et~al.}{2011}]{Nan2011}
{Nan} R.,  et~al., 2011, \mn@doi [International Journal of Modern Physics D]
  {10.1142/S0218271811019335}, \href
  {https://ui.adsabs.harvard.edu/abs/2011IJMPD..20..989N} {20, 989}

\bibitem[\protect\citeauthoryear{{Sadler} et~al.,}{{Sadler}
  et~al.}{2007}]{Sadler2007}
{Sadler} E.~M.,  et~al., 2007, \mn@doi [\mnras]
  {10.1111/j.1365-2966.2007.12231.x}, \href
  {https://ui.adsabs.harvard.edu/abs/2007MNRAS.381..211S} {381, 211}

\bibitem[\protect\citeauthoryear{{Sadler} et~al.,}{{Sadler}
  et~al.}{2020}]{Sadler2020}
{Sadler} E.~M.,  et~al., 2020, \mn@doi [\mnras] {10.1093/mnras/staa2390}, \href
  {https://ui.adsabs.harvard.edu/abs/2020MNRAS.499.4293S} {499, 4293}

\bibitem[\protect\citeauthoryear{{Saintonge}}{{Saintonge}}{2007}]{Saintonge2007}
{Saintonge} A.,  2007, \mn@doi [\aj] {10.1086/513515}, \href
  {https://ui.adsabs.harvard.edu/abs/2007AJ....133.2087S} {133, 2087}

\bibitem[\protect\citeauthoryear{Smith, Dunning, Smart, Shaw, Mackay, Bowen  \&
  Hayman}{Smith et~al.}{2017}]{Smith2017}
Smith S.~L.,  Dunning A.,  Smart K.~W.,  Shaw R.,  Mackay S.,  Bowen M.,
  Hayman D.,  2017, in 2017 IEEE International Symposium on Antennas and
  Propagation \& USNC/URSI National Radio Science Meeting. pp 2137--2138,
  \mn@doi{10.1109/APUSNCURSINRSM.2017.8073111}

\bibitem[\protect\citeauthoryear{{Srianand}, {Gupta}, {Petitjean}, {Noterdaeme}
   \& {Saikia}}{{Srianand} et~al.}{2008}]{Srianand2008}
{Srianand} R.,  {Gupta} N.,  {Petitjean} P.,  {Noterdaeme} P.,   {Saikia}
  D.~J.,  2008, \mn@doi [\mnras] {10.1111/j.1745-3933.2008.00558.x}, \href
  {https://ui.adsabs.harvard.edu/abs/2008MNRAS.391L..69S} {391, L69}

\bibitem[\protect\citeauthoryear{{Uson}, {Bagri}  \& {Cornwell}}{{Uson}
  et~al.}{1991}]{Uson1991}
{Uson} J.~M.,  {Bagri} D.~S.,   {Cornwell} T.~J.,  1991, \mn@doi [\prl]
  {10.1103/PhysRevLett.67.3328}, \href
  {https://ui.adsabs.harvard.edu/abs/1991PhRvL..67.3328U} {67, 3328}

\bibitem[\protect\citeauthoryear{{Wegner}, {Colless}, {Saglia}, {McMahan},
  {Davies}, {Burstein}  \& {Baggley}}{{Wegner} et~al.}{1999}]{Wegner1999}
{Wegner} G.,  {Colless} M.,  {Saglia} R.~P.,  {McMahan} R.~K.,  {Davies} R.~L.,
   {Burstein} D.,   {Baggley} G.,  1999, \mn@doi [\mnras]
  {10.1046/j.1365-8711.1999.02339.x}, \href
  {https://ui.adsabs.harvard.edu/abs/1999MNRAS.305..259W} {305, 259}

\bibitem[\protect\citeauthoryear{{Wu}, {Haynes}, {Giovanelli}, {Zhu}  \&
  {Chen}}{{Wu} et~al.}{2015}]{Wu2015}
{Wu} Z.,  {Haynes} M.~P.,  {Giovanelli} R.,  {Zhu} M.,   {Chen} R.-r.,  2015,
  \mn@doi [Acta Astronomica Sinica (in Chinese)]
  {10.15940/j.cnki.0001-5245.2015.02.003}, 56, 112

\bibitem[\protect\citeauthoryear{{Yu}, {Pen}, {Zhang}, {Li}  \& {Chen}}{{Yu}
  et~al.}{2017}]{Yu2017}
{Yu} H.-R.,  {Pen} U.-L.,  {Zhang} T.-J.,  {Li} D.,   {Chen} X.,  2017, \mn@doi
  [Research in Astronomy and Astrophysics] {10.1088/1674-4527/17/6/49}, \href
  {https://ui.adsabs.harvard.edu/abs/2017RAA....17...49Y} {17, 049}

\bibitem[\protect\citeauthoryear{{Zhang}, {Xiao}, {Yu}, {Luo}, {Yang}  \&
  {Zhu}}{{Zhang} et~al.}{2019}]{Zhang2019}
{Zhang} B.,  {Xiao} J.,  {Yu} C.,  {Luo} Q.,  {Yang} Z.,   {Zhu} M.,  2019, in
  Proceedings of the 2018 Radio Astronomy Forum: FAST-MeerKAT and SKA
  Pathfinders Synergies. Huazhong University of Science and Technology Press -
  Wuhan, p.~174

\bibitem[\protect\citeauthoryear{{Zheng}, {Li}, {Sadler}, {Allison}  \&
  {Tang}}{{Zheng} et~al.}{2020}]{Zheng2020}
{Zheng} Z.,  {Li} D.,  {Sadler} E.~M.,  {Allison} J.~R.,   {Tang} N.,  2020,
  \mn@doi [\mnras] {10.1093/mnras/staa3033}, \href
  {https://ui.adsabs.harvard.edu/abs/2020MNRAS.499.3085Z} {499, 3085}

\bibitem[\protect\citeauthoryear{{Zwaan}, {van der Hulst}, {Briggs},
  {Verheijen}  \& {Ryan-Weber}}{{Zwaan} et~al.}{2005}]{Zwaan2005}
{Zwaan} M.~A.,  {van der Hulst} J.~M.,  {Briggs} F.~H.,  {Verheijen} M.~A.~W.,
   {Ryan-Weber} E.~V.,  2005, \mn@doi [\mnras]
  {10.1111/j.1365-2966.2005.09698.x}, \href
  {https://ui.adsabs.harvard.edu/abs/2005MNRAS.364.1467Z} {364, 1467}

\makeatother
\end{thebibliography}








\bsp	
\label{lastpage}
\end{document}